\documentclass[sigconf]{acmart}
\AtBeginDocument{%
  }

\setcopyright{acmlicensed}
\copyrightyear{2026}
\acmYear{2026}
\acmDOI{XXXXXXX.XXXXXXX}
\acmConference[Conference ICSE '26]{ICSE '26: 48th International Conference on Software Engineering}{April 12-18, 2026}{Rio de Janeiro, Brazil}
\acmISBN{978-1-4503-XXXX-X/2026/06}
\usepackage{xspace}
\usepackage{enumitem}
\usepackage{multirow,multicol}
\usepackage{algorithm}
\usepackage{algorithmic}
\usepackage{listings}
\usepackage{amsmath}
\usepackage{makecell}
\usepackage[most]{tcolorbox}
\usepackage{array}

\newtcolorbox[auto counter,number within=section]{promptfigure}[2][]{
    colback=blue!5,
    colframe=blue!75!black,
    fonttitle=\bfseries,
    title=Prompt Template \thetcbcounter: #2,
    float=htbp,
    every float=\centering,
    #1
}

\newtcolorbox[auto counter,number within=section]{dialogfigure}[2][]{
    colback=gray!5,
    colframe=gray!75!black,
    fonttitle=\bfseries,
    title=Completion Example \thetcbcounter: #2,
    float=htbp,
    every float=\centering,
    #1
}

\newcommand{\modelname}{GRACE\xspace}
\newcommand{\stitle}[1]{\vspace{0.8mm} \noindent {\bf #1}}

\begin{document}

\title{GRACE: Graph-Guided Repository-Aware Code Completion through Hierarchical Code Fusion}



\author{Xingliang Wang}
\affiliation{%
  \institution{Zhejiang University}
  \city{Hangzhou}
  \state{Zhejiang}
  \country{China}
}
\email{wangxingliang@zju.edu.cn}

\author{Baoyi Wang}
\affiliation{%
  \institution{Zhejiang University}
  \city{Hangzhou}
  \state{Zhejiang}
  \country{China}
}
\email{wangbaoyi@zju.edu.cn}

\author{Haoran Xu}
\affiliation{%
  \institution{Zhejiang University}
  \city{Hangzhou}
  \state{Zhejiang}
  \country{China}
}
\email{haoran.x@zju.edu.cn}

\author{Chen Zhi}
\authornote{Corresponding author}
\affiliation{%
  \institution{Zhejiang University}
  \city{Hangzhou}
  \state{Zhejiang}
  \country{China}
}
\email{zjuzhichen@zju.edu.cn}

\author{Junxiao Han}
\affiliation{%
  \institution{Hangzhou City University}
  \city{Hangzhou}
  \state{Zhejiang}
  \country{China}
}
\email{hanjx@hzcu.edu.cn}

\author{Xinkui Zhao}
\affiliation{%
  \institution{Zhejiang University}
  \city{Hangzhou}
  \state{Zhejiang}
  \country{China}
}
\email{zhaoxinkui@zju.edu.cn}

\author{Jianwei Yin}
\affiliation{%
  \institution{Zhejiang University}
  \city{Hangzhou}
  \state{Zhejiang}
  \country{China}
}
\email{zjuyjw@zju.edu.cn}

\author{Shuiguang Deng}
\affiliation{%
  \institution{Zhejiang University}
  \city{Hangzhou}
  \state{Zhejiang}
  \country{China}
}
\email{dengsg@zju.edu.cn}

\renewcommand{\shortauthors}{Trovato et al.}

\begin{abstract}
Large language models (LLMs) excel in localized code completion but struggle with repository-level tasks due to limited context windows and complex semantic and structural dependencies across codebases. While Retrieval-Augmented Generation (RAG) mitigates context scarcity by retrieving relevant code snippets, current approaches face significant limitations. They overly rely on textual similarity for retrieval, neglecting structural relationships such as call chains and inheritance hierarchies, and lose critical structural information by naively concatenating retrieved snippets into text sequences for LLM input.
To address these shortcomings, we propose GRACE, a novel framework that reimagines code repositories as hierarchical, semantically rich graph databases, fundamentally advancing repository-level code completion. GRACE constructs a multi-level, multi-semantic code graph that unifies file structures, abstract syntax trees, function call graphs, class hierarchies, and data flow graphs to capture both static and dynamic code semantics. For retrieval, GRACE employs a Hybrid Graph Retriever that integrates graph neural network-based structural similarity with textual retrieval, refined by a graph attention network-based re-ranker to prioritize topologically relevant subgraphs. To enhance context, GRACE introduces a structural fusion mechanism that merges retrieved subgraphs with the local code context and serializes the unified graph into LLM-readable text with explicit structural markers, preserving essential dependencies like function calls and inheritance.
This work provides the first systematic analysis of graph structures in repository-level code completion, introducing a novel multi-semantic graph representation, a structure-aware hybrid retrieval and reranking strategy, and an innovative graph fusion technique that prevents structural information loss. Extensive experiments on public repository-level benchmarks demonstrate that GRACE significantly outperforms state-of-the-art methods across all metrics. Using DeepSeek-V3 as the backbone LLM, GRACE surpasses the strongest graph-based RAG baselines by 8.19\% EM and 7.51\% ES points on every dataset. The code is available at https://anonymous.4open.science/r/grace\_icse-C3D5.

\end{abstract}

\ccsdesc[500]{Do Not Use This Code~Generate the Correct Terms for Your Paper}
\ccsdesc[300]{Do Not Use This Code~Generate the Correct Terms for Your Paper}
\ccsdesc{Do Not Use This Code~Generate the Correct Terms for Your Paper}
\ccsdesc[100]{Do Not Use This Code~Generate the Correct Terms for Your Paper}

\keywords{Code Completion, Code Generation, Large Language Model, Graph, RAG}


\maketitle

\section{Introduction}
\label{sec:intro}
In recent years, large language models\cite{openai2024gpt4o,grattafiori2024llama,nijkamp2023codegen2,roziere2023code,allal2023santacoder,zheng2023codegeex,christopoulou2022pangu,guo2024deepseek,chowdhery2023palm} have demonstrated remarkable capabilities in function-level or file-level code completion tasks\cite{zheng2023survey,zhang2023unifying,zan2022large,yadavalli2023rlpg,bairi2024codeplan}. However, real-world software development typically involves large-scale, multi-module code repositories whose complexity far exceeds that of individual files. In this context, code completion faces significant challenges: (1) the vast scale of codebases exceeds the limited context window of LLMs\cite{shi2023large}; (2) intricate semantic and structural dependencies (e.g., inheritance, implementation, call chains, data flow) exist between code entities (classes, functions, variables). These factors lead to a substantial degradation in performance when LLMs are directly applied to repository-level code completion\cite{tao2025code,tang2023domain,zan2022cert,zhang2023repocoder}.

To address these issues, the Retrieval-Augmented Generation (RAG) paradigm has gradually become the mainstream solution\cite{lewis2020retrieval,agrawal2023guiding,zhang2023repocoder,parvez2021retrieval,liao2023context}. Although RAG partially mitigates the context limitation problem, existing RAG techniques for repository-level code completion suffer from two critical limitations.

\begin{figure*}[t]
    \centering
    \includegraphics[width=\textwidth]{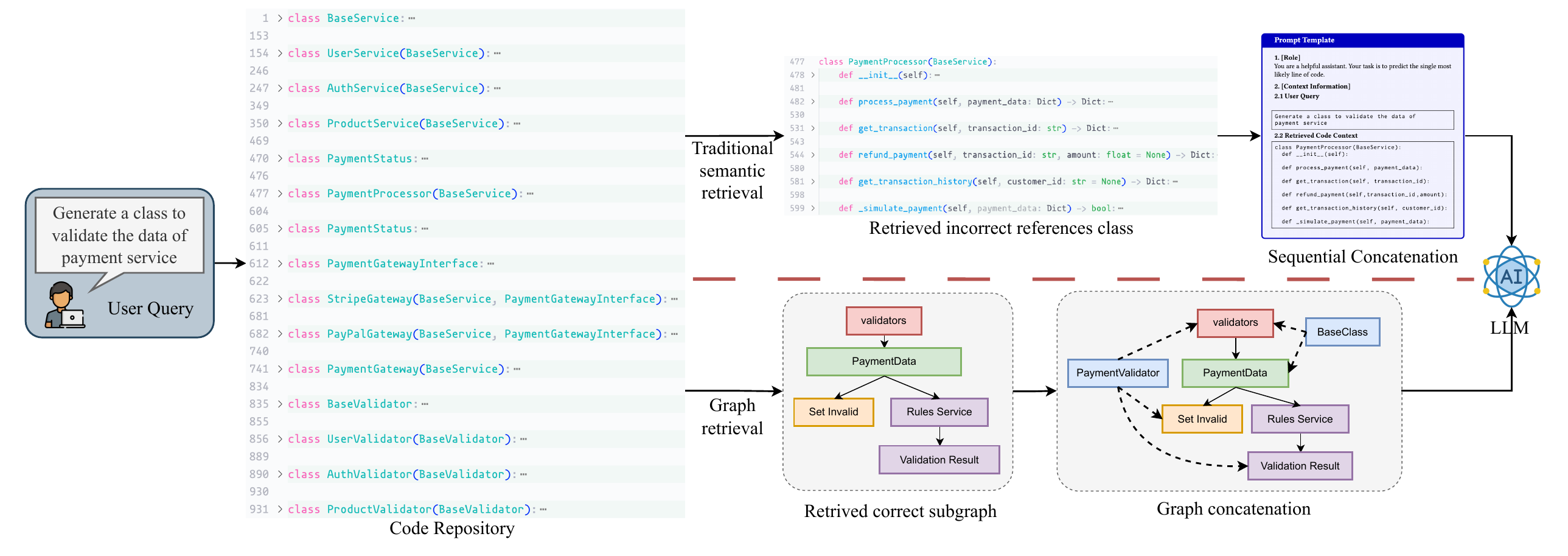}
    \caption{Comparison of standard RAG-based code completion with GRACE. Conventional RAG retrieves code solely by semantic similarity and fuses it via naive concatenation, ignoring structural relations and often producing inaccurate completions.}
    \label{fig:introduction}
\end{figure*}

First, most existing studies\cite{zhang2023repocoder,zan2022language,li2023acecoder} rely on semantic similarity to retrieve relevant code while neglecting structural information as shown in Figure \ref{fig:introduction}. Current RAG methods typically treat code as plain text sequences and primarily rely on traditional text retrieval techniques such as BM25\cite{robertson2009probabilistic}, embedding-based similarity, or TF-IDF. This approach struggles to capture the rich structural information inherent in code\cite{peng2024graph}, including class inheritance hierarchies, interface implementation patterns, cross-module call dependencies, and complex dataflow paths. While a few studies attempt to incorporate simple graph structures\cite{liu2020multi,clement2021long} like Abstract Syntax Trees (ASTs)\cite{li2017code,kim2021code}, their utilization of code structural information remains rudimentary, failing to adequately exploit deep semantic relationships.
Second, the utilization of retrieved results is limited to text-level concatenation . Existing methods\cite{liu2024graphcoder,liang2024repofuse,phan2024repohyper} commonly adopt simple sequential concatenation strategies as shown in Figure \ref{fig:introduction}, feeding the concatenation of retrieved code snippets into the LLM. This approach entirely ignores potential structural relationships between the retrieved code snippets and the incomplete code, such as variable definition-use chains, method override inheritance, and cross-snippet call dependencies. The absence of this information severely constrains the LLM's ability to understand and leverage the retrieved context.

To address these issues, we propose a fundamental paradigm shift: modeling the entire code repository as a hierarchical, semantically-rich code graph database and designing a novel graph-based RAG pipeline on this foundation. The core of this approach lies in leveraging the graph structure's ability to express code elements and their rich semantic relationships. However, applying this concept to repository-level code completion RAG pipelines faces several core challenges:
\begin{enumerate}
    \item Graph Construction: How to transform a large and complex code repository into a unified graph structure that can both comprehensively represent code entities (classes, methods, variables, etc.) and precisely characterize the complex relationships between entities (calls, inheritance, data flows, etc.)?
    \item Graph Retrieval: How to design a high-performance retrieval strategy on the constructed code graph that captures both structural similarity and semantic relevance to provide optimal contextual information for the code completion task?
    \item Graph Augmentation: How to effectively fuse the retrieved graph structure information with the code-to-be-completed, fully utilizing the structural associations between them to enhance the LLM's code understanding and generation capabilities?
\end{enumerate}

To systematically address these challenges, we propose the \textbf{G}raph-Guided \textbf{R}epository-\textbf{A}ware \textbf{C}ode Compl\textbf{e}tion through Hierarchical Code Fusion (GRACE) framework.
In the graph construction phase, we extract and build a multi-level, multi-semantic hybrid code graph from the code repository. We categorize the graph structures into three levels: repo-level, module-level, and function-level. We capture semantic relationships of different granularities at each level. Through organic fusion between levels, we construct a unified code graph capable of expressing both static structural features and dynamic behavioral patterns of code.
In the graph retrieval phase, we propose a Hybrid Graph Retriever (HGR) that innovatively combines GNN-based structural retrieval with traditional text semantic retrieval. Structural retrieval identifies code snippets with similar topological patterns (such as similar call chains or inheritance structures) through subgraph matching, while text retrieval captures semantic-level relevance. We further design a graph-aware reranking mechanism to refine the preliminary retrieval results, ensuring the prioritization of the most relevant structured context.
In the graph enhancement phase, we propose a Graph Fusion and Enhancement mechanism that fuses the optimal retrieved code subgraph with the graph structure of the code-to-be-completed both semantically and structurally, forming a more informationally complete enhanced context graph, thereby significantly improving the LLM's understanding of code structure.

The main contributions of this paper are as follows:
\begin{itemize}[leftmargin=0.4cm] 
    \item We are the first to systematically propose a multi-level, multi-semantic code graph construction method for repository-level code completion, effectively addressing the insufficient utilization of code structural information in existing methods.
    \item We design and implement a hybrid graph retrieval strategy that integrates structural and semantic retrieval, combined with a graph-aware reranking mechanism, significantly improving the quality and relevance of retrieved context.
    \item We innovatively propose a graph fusion enhancement mechanism that, for the first time, focuses on and utilizes the structural associations between retrieved code and code-to-be-completed, effectively solving the structural information loss problem in traditional RAG methods.
    \item We conduct comprehensive experimental evaluations on multiple public repository-level code completion benchmark datasets, with results showing that the \modelname framework significantly outperforms existing state-of-the-art methods. Using DeepSeek-V3 as the backbone LLM, GRACE surpasses the strongest graph-based RAG baselines by 8.19\% EM and 7.51\% ES points on four datasets.
\end{itemize}
\section{Methodology}

\begin{figure*}[t]
  \centering
  \includegraphics[width=\linewidth]{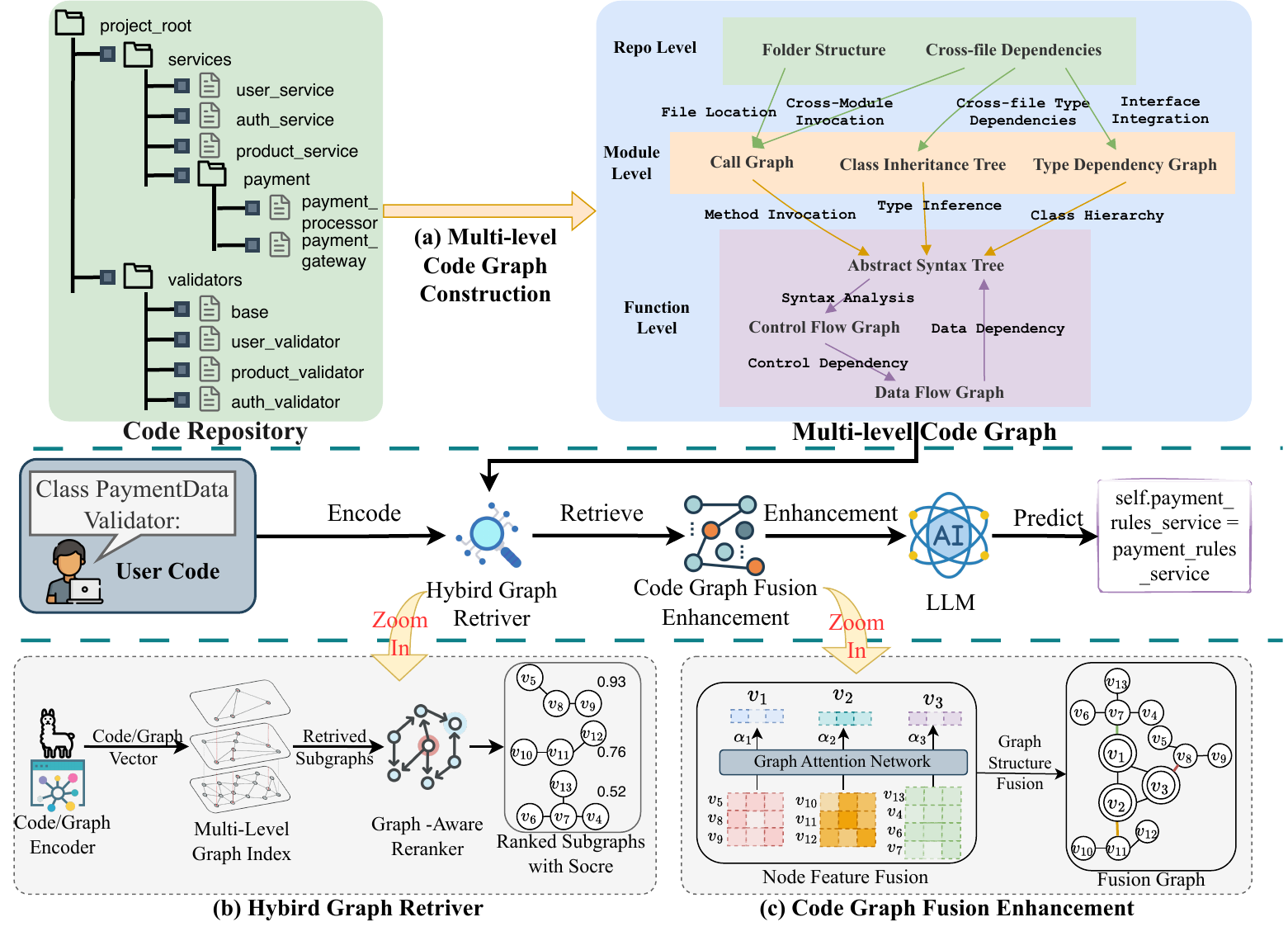}
  \caption{Framework of \modelname, which consists of three main components: (a) multi-level code graph construction, (b) hybrid graph retriever, and (c) code graph fusion enhancement.}
  \label{fig:framework}
  \end{figure*}

\subsection{Multi-level Code Graph Construction}

In repository-level code completion tasks, the hierarchical structure and multi-dimensional semantic relationships of code are crucial for understanding code context. Code repositories inherently exhibit hierarchical and multi-relational characteristics. From project file structures to function statements, code entities at different granularities (such as packages, modules, classes, and functions) carry semantic information at different levels. Traditional graph representations typically only use simple file structures and ASTs, failing to comprehensively capture cross-level semantic dependencies, resulting in the loss of critical context during the retrieval-augmented process. To fully capture this rich structured information, we propose a multi-level, multi-semantic code graph construction method that models the entire code repository as a unified hierarchical graph structure through a three-level abstraction and structured connection mechanism.

\subsubsection{Multi-level Code Graph Structure Design}

We divide the repository-level code graph structure into three levels, with each level focusing on code semantics at different granularities:

\begin{table}[t]
\centering
\caption{Hierarchical graph representations leveraged by our RAG framework}
\label{tab:graph_reps}
\begin{tabular}{@{}m{0.15\linewidth} m{0.37\linewidth} m{0.42\linewidth}@{}}
\toprule
\textbf{Level} & \textbf{Graph Type} & \textbf{Semantics Captured} \\
\midrule
\multirowcell{4}{Repository} 
& Folder Structure Graph & Directory hierarchy; nodes are folders/files, edges denote containment. \\ \cmidrule(lr){2-3}
& Cross-file Dep. Tree & Import and reference relations between files, reflecting module coupling. \\
\midrule
\multirowcell{6}{Module} 
& Function Call Graph & Call relations among functions. \\ \cmidrule(lr){2-3}
& Type Dep. Graph & Dependencies between type definitions and their usages. \\ \cmidrule(lr){2-3}
& Class Inheritance Graph & Inheritance, implementation, and composition relations among classes. \\
\midrule
\multirowcell{5}{Function} 
& Abstract Syntax Tree & Syntactic structure of code. \\ \cmidrule(lr){2-3}
& Control Flow Graph & Possible execution paths of the program. \\ \cmidrule(lr){2-3}
& Data Flow Graph & Definition–use relations of variables. \\
\bottomrule
\end{tabular}
\end{table}

\subsubsection{Construction of Unified Hierarchical Graph}

To achieve unified representation and efficient retrieval across levels, we integrate the extracted multi-granularity graph structures into a unified multi-relational hierarchical graph $\mathcal{G} = (\mathcal{V}, \mathcal{E}, \mathcal{T}_v, \mathcal{T}_e)$, where $\mathcal{V}$ represents the set of nodes across all levels, $\mathcal{E}$ represents the set of edges including both intra-level and cross-level edges, $\mathcal{T}_v$ represents the set of node types, including files, classes, functions, statements, etc.; $\mathcal{T}_e$ represents the set of edge types, including calls, inheritance, definitions, uses, etc.

To avoid naming conflicts between different levels, we assign a globally unique identifier (UUID) to each node and maintain a mapping table that records the original information of nodes, including the graph type they belong to (e.g., $graph\_type = \text{"call\_graph"}$), node type, and attribute information. Node attributes include code text, location information (file path, line number), semantic type (such as variable types, function signatures), and structural features (such as cyclomatic complexity, nesting depth). Edge attributes include edge type, weight (such as call frequency, dependency strength), and context information (such as call parameters, conditional constraints). These attribute information not only preserve the original semantics of the code but also provide rich features for subsequent graph retrieval and matching.

To bridge the three granularities, we introduce a set of carefully designed cross-level edges (see Figure \ref{fig:framework} (a)).  
At the repository–module boundary, every file node is linked to the functions it declares, thereby localising definition sites; file-level dependencies are further propagated to the corresponding inter-function calls, while cross-file type usages and inheritance relations are mapped to explicit ``type-reference'' and ``interface-inheritance'' edges.  
Moving down to the module-function boundary, each function in the call graph is anchored to the root of its abstract syntax tree, and variable types are aligned with data-flow vertices so that type information constrains the data-flow analysis.
Within a single function, AST nodes are connected to the control-flow graph to reflect the influence of syntax on execution order; control-flow vertices are then tied to data-flow vertices, capturing how branching and looping conditions restrict the propagation of values.  
These cross-level connections weave repository, module, and function views into one heterogeneous graph, providing a unified and semantically rich backdrop for subsequent retrieval and code completion.

Through the above multi-level graph construction method, we transform complex code repositories into a structured, hierarchical code knowledge base, providing a solid foundation for subsequent graph retrieval and code completion tasks.

\subsection{Hybrid Graph Retriever}

After constructing the multi-level code graph, the key challenge lies in retrieving the most relevant context from this large-scale graph database to assist code completion. In repository-level code completion scenarios, developers often need to understand not only what a piece of code does (semantic information) but also how it interacts with other components (structural information). For instance, when completing a method call, knowing similar function implementations helps, but understanding the inheritance hierarchy, method overrides, and calling patterns provides crucial constraints for generating correct completions. Traditional retrieval methods that rely solely on textual similarity miss these structural dependencies, leading to semantically plausible but structurally incorrect suggestions. To address this fundamental limitation, we propose the Hybrid Graph Retriever (HGR) (see Figure \ref{fig:framework} (b)), which synergistically combines semantic retrieval for capturing functional similarity and graph-based structural retrieval for understanding code relationships, thereby providing comprehensive context for more accurate code completion.

\subsubsection{Dual-Path Encoding Mechanism}

The Hybrid Graph Retriever employs a dual-path encoding strategy that captures both the semantics and structure of code, each serving distinct but complementary roles in code completion.

\textbf{Semantic Encoding Path}: For semantic information in code, we employ the pre-trained Code embedding model (codet5p-110m-embedding) to encode code snippets. This model, pre-trained on large-scale mixed code-natural language corpora, effectively captures functional semantics, variable naming conventions, and comment information. For each code snippet $c$, the semantic encoder generates a corresponding semantic vector $\mathbf{v}_c \in \mathbb{R}^{d_s}$:
\begin{equation}
   \mathbf{v}_c \;=\;
   \mathrm{Embed}_{\mathrm{CodeT5p}}\!\left(c\right)\in\mathbb{R}^{d_s},
 \end{equation}
where $d_s$ denotes the semantic embedding dimension. The generated semantic vectors are indexed using Hierarchical Navigable Small World (HNSW) graphs, supporting efficient approximate nearest neighbor search.
Notably, the hierarchical nature of HNSW aligns well with our multi-level code graph structure—upper layers provide fast navigation across different modules while lower layers enable precise retrieval within specific code regions.

\textbf{Structural Encoding Path}: 
For structural information in code, we encode this structural information through a combination of node embeddings and Laplacian positional encoding: First, we utilize the code language model to encode the code content of each node $v\!\in\!\mathcal{V}$ in the graph, obtaining node embeddings $\mathbf{v}_c\in\mathbb{R}^{d_1}$. Second, to capture the global positional information of nodes within the graph, we compute the normalized Laplacian matrix:

$$
\mathbf{L}=\mathbf{I}-\mathbf{D}^{-1/2}\mathbf{A}\mathbf{D}^{-1/2},
$$

where $\mathbf{A}$ is the adjacency matrix and $\mathbf{D}$ is the degree matrix. We take the eigenvectors corresponding to the first $d_2$ smallest non-zero eigenvalues of $\mathbf{L}$ as the positional encoding $\mathbf{v}_s \in \mathbb{R}^{d_2}$ for each node. The final graph structural representation is obtained through concatenation:

\begin{equation}
   \mathbf{h}_v \;=\; [\,\mathbf{v}_c \,;\, \mathbf{v}_s\,]\in\mathbb{R}^{d_1+d_2},
 \end{equation}

This dual encoding mechanism preserves both the local semantic information of nodes and encodes their positional features within the global graph structure.

\subsubsection{Hybrid Retrieval Strategy}

Based on the dual-path encoding, the Hybrid Graph Retriever executes semantic and structural retrieval in parallel, fully leveraging the complementary advantages of both retrieval paradigms.

\textbf{Semantic Similarity Retrieval}: Given a query code snippet $q$, we first generate the query vector $\mathbf{v}_q$ through the semantic encoder. Using the HNSW index, we retrieve the Top-$k_s$ most semantically relevant code snippets based on cosine similarity:

$$\text{SemSim}(q, c) = \frac{\mathbf{v}_q \cdot \mathbf{v}_c}{\|\mathbf{v}_q\| \|\mathbf{v}_c\|}$$

Semantic retrieval effectively captures code snippets with similar functionality but different implementations, demonstrating good generalization capability.

\textbf{Structural Similarity Retrieval}: In parallel, we parse the query code into a local graph structure $G_q$ and generate graph embeddings $\mathbf{h}_{G_q}$ through the structural encoder. We retrieve the Top-$k_g$ most structurally matching subgraphs from the graph vector index:

$$\text{StructSim}(G_q, G_c) = \frac{\mathbf{h}_{G_q} \cdot \mathbf{h}_{G_c}}{\|\mathbf{h}_{G_q}\| \|\mathbf{h}_{G_c}\|}$$

Structural retrieval identifies code snippets with similar calling patterns, inheritance relationships, or control flow structures, which is crucial for understanding the deep semantics of code.

\subsubsection{Graph-Aware Reranking Mechanism}

The candidate sets retrieved from semantic and structural dimensions may overlap or contain complementary information. To optimize the final retrieval results, we design a graph-aware reranking mechanism.
We first merge the Top-$k_s$ candidate set $\mathcal{C}_{\text{sem}}$ from semantic retrieval with the Top-$k_g$ candidate set $\mathcal{C}_{\text{struct}}$ from structural retrieval, obtaining the combined candidate set $\mathcal{C} = \mathcal{C}_{\text{sem}} \cup \mathcal{C}_{\text{struct}}$.
For each candidate $c_i \in \mathcal{C}$ in the merged set, we compute its relevance score:
$$\text{Score}(c_i) = \alpha \cdot \text{SemSim}(q, c_i) + (1-\alpha) \cdot \text{StructSim}(G_q, G_{c_i}),$$
where $\alpha \in [0,1]$ is a balancing parameter controlling the relative importance of semantic and structural similarity. In practice, we employ an adaptive attention mechanism to dynamically adjust $\alpha$:
$$\alpha = \sigma(W_{\alpha}[\mathbf{v}_q; \mathbf{h}_{G_q}] + b_{\alpha}),$$
where $W_{\alpha}$ and $b_{\alpha}$ are learnable parameters, and $\sigma$ is the sigmoid function. This adaptive mechanism automatically adjusts the weights of semantic and structural information based on query.

To avoid overly homogeneous retrieval results, we introduce diversity constraints. Through the Maximal Marginal Relevance (MMR) algorithm, we enhance result diversity while maintaining relevance:
$$\text{MMR}(c_i) = \lambda \cdot \text{Score}(c_i) - (1-\lambda) \cdot \max_{c_j \in S} \text{Sim}(c_i, c_j),$$
where $S$ is the set of already selected candidates code snippets, and $\lambda$ controls the balance between relevance and diversity.

\subsection{Code Graph Fusion Enhancement}
After obtaining high-quality retrieval results, effectively leveraging this structured information becomes a critical factor determining code completion quality. Traditional RAG approaches simply concatenate retrieved code snippets with the code to be completed, ignoring potential structural associations between them. This treatment leads to the loss of valuable structural information (such as cross-snippet call dependencies, shared type constraints, and similar control flow patterns) before being fed into the LLM.

To address this limitation, we propose a graph fusion enhancement mechanism (see Figure \ref{fig:framework} (c)). The core idea is to construct a query graph from the user's incomplete code and then perform deep fusion between multiple retrieved code subgraphs and the query graph across two dimensions: node features and graph structure. This creates a unified enhanced context graph that provides the LLM with more complete and structured contextual information.

\subsubsection{Fusion Mechanism Design}

The query graph $G_q = (V_q, E_q)$ is an AST graph structure constructed from the user's current incomplete code snippet. This graph contains local syntactic structures, variable definitions, function calls, and other information from the code to be completed. However, due to the code's incompleteness, it often lacks necessary contextual dependencies.

Rich structural associations exist between retrieved code subgraphs and the query graph, which are invaluable for code completion tasks. Similar function implementations in retrieved subgraphs can provide implementation patterns and parameter constraints for incomplete functions in the query graph, offering crucial semantic completion cues. Additionally, complete call paths in retrieved subgraphs can provide contextual validation for function calls in the query graph, ensuring call chain completeness.

Our dual-dimensional fusion approach leverages these associations through complementary mechanisms. Node feature fusion aligns query graph nodes with retrieved graph nodes in semantic space, enabling the LLM to identify functionally similar code entities and generate semantically consistent completions. Simultaneously, graph structure fusion preserves and extends dependencies between code components, enabling the LLM to understand call constraints and data flow, thereby generating structurally correct completions.

\subsubsection{Node Feature Fusion}

Node feature fusion aims to establish correspondences between query graphs and retrieved graphs in semantic space. Given a query graph $G_q = (V_q, E_q)$ and $k$ retrieved subgraphs $\{G_1, G_2, ..., G_k\}$, we first obtain node representations for each graph through graph neural network encoders:

\begin{align}
\mathbf{H}_q &= \text{GNN}(G_q) \in \mathbb{R}^{|V_q| \times d} \\
\mathbf{H}_i &= \text{GNN}(G_i) \in \mathbb{R}^{|V_i| \times d}, \quad i = 1, 2, ..., k
\end{align}

To integrate multiple retrieval results, we perform weighted aggregation of retrieved graphs based on retrieval relevance scores:
\begin{equation}
\mathbf{H}_r = \sum_{i=1}^{k} w_i \cdot \mathbf{H}_i
\end{equation}

where weights $w_i$ are obtained through normalization of retrieval relevance scores. Subsequently, we compute cross-attention between query graph nodes and aggregated retrieved graph nodes:
\begin{equation}
\mathbf{A} = \text{softmax}\left(\frac{\mathbf{H}_q \mathbf{H}_r^T}{\sqrt{d}}\right) \in \mathbb{R}^{|V_q| \times |V_r|}
\end{equation}

This attention mechanism enables the identification of functionally similar code entities (e.g., functions with identical functionality) and provision of semantic alignment foundation for subsequent structural fusion.

\subsubsection{Graph Structure Fusion}

Building upon semantic alignment from node features, we further establish connections between query graphs and retrieved graphs at the graph structure level. For node pairs $(v_q, v_r)$ with attention weights exceeding threshold $\theta$, we create cross-graph edges in the fusion graph:

\begin{equation}
E_{\text{fusion}} = E_q \cup E_r \cup \{(v_q, v_r) | \mathbf{A}_{v_q,v_r} > \theta\}
\end{equation}

To ensure fusion graph quality and code semantic correctness, we introduce several essential constraints. First, we enforce type consistency by only connecting type-compatible nodes, such as function calls with function definitions or variables of the same type. Second, we preserve critical paths and dependency relationships from original graphs to avoid disrupting code's inherent logic. Finally, we eliminate redundancy by merging semantically equivalent nodes to avoid information duplication and confusion.

\subsubsection{Graph-to-Text Serialization}

In order to input the structured enhanced context graph into the LLM, we employ a semantic graph serialization method that preserves both structural and semantic information. Our serialization template explicitly expresses node types and relationship types, providing structural clarity that facilitates LLM comprehension. The representation maintains semantic richness by preserving property information and weight information, while the natural language format ensures comprehensibility for effective LLM processing.
The final prompt template combines the serialized graph structure with the original code context:

\begin{promptfigure}{}

   \small
   \textbf{1. [Role]} \\
   You are a world-class AI code completion expert. Your purpose is to help developers write code faster and more accurately. You will be given the user's current code context and a relevant code knowledge subgraph retrieved from the entire codebase. Your task is to predict the single most likely line of code to complete at the cursor position.
   
   \vspace{0.5em}
   \textbf{2. [Context Information]} \\[0.2em]
   \textbf{2.1 User's Current Code Context} \\
   {\footnotesize
   \begin{tabular}{@{}ll@{}}

   \end{tabular}}
   \begin{lstlisting}[language=,numbers=none,frame=single,basicstyle=\ttfamily\footnotesize]
   repo name: {repo_name}
   File Path: {current_file_path}
   {code_before_cursor}
   \end{lstlisting}
   
   \textbf{2.2 Retrieved Code Context}
   \begin{lstlisting}[language=,numbers=none,frame=single,basicstyle=\ttfamily\footnotesize]
   {code_context}
   \end{lstlisting}
   
   \textbf{2.3 Retrieved Code Knowledge Graph}
   \begin{lstlisting}[language=,numbers=none,frame=single,basicstyle=\ttfamily\footnotesize]
   {graph_context}
   \end{lstlisting}
   
   \vspace{0.5em}
   \textbf{2. [Task Instruction]} \\
   Analyze the user's current code context to understand their immediate goal. Examine the provided code knowledge subgraph, focusing on class definitions, function signatures, and usage patterns. Synthesize this information to infer the most logical next line of code. Output \emph{exactly one} line of code for insertion and a concise explanation referencing specific nodes from the subgraph.
   
   \vspace{0.5em}
   \textbf{3. [Output Format]} \\
   Return a single valid JSON object:
   \begin{lstlisting}[numbers=none,frame=single,basicstyle=\ttfamily\footnotesize]
   {
     "completed_code": "The suggested code.",
     "explanation": "Brief rationale referencing 
     subgraph nodes.",
     "confidence_score": 0.87,
     "referenced_nodes": [
       "node_id_of_relevant_function",
       "node_id_of_relevant_class"
     ]
   }
   \end{lstlisting}
   
   \vspace{0.5em}
   \textbf{4. [Constraints and Rules]} \\
   • The completed\_code must contain exactly one line. \\
   • Explanations must remain concise and grounded in the subgraph. \\
   • If uncertain, lower the confidence\_score; if impossible, return an empty code line with confidence 0.0.
   
   \end{promptfigure}

Through this graph fusion enhancement mechanism, we provide the LLM with rich context that encompasses both semantic similarity and structural consistency, thereby significantly improving the accuracy and executability of code completion.

\subsection{Algorithm Implementation}
The complete algorithmic flow for \modelname is presented in Algorithm~\ref{alg:grace_pipeline}. The process can be decomposed into five distinct phases: (1) Query Graph Construction, (2) Hybrid Graph Retrieval, (3) Graph Fusion Enhancement, (4) Prompt Serialization, and (5) LLM Inference.

\textbf{Phase 1: Query Graph Construction.}
The process starts with the user’s incomplete code snippet $\mathcal{C}$, which is parsed into an Abstract Syntax Tree (AST). From this AST, we build the query graph $G_q = (V_q, E_q)$.

\textbf{Phase 2: Hybrid Graph Retrieval.}
Our retrieval mechanism adopts a hybrid approach that combines both semantic and structural similarity, aiming to leverage the complementary strengths of the two retrieval methods.

For the \textbf{semantic path}, we first encode the query graph $G_q$ into a high-dimensional vector embedding $\mathbf{v}_q$ using the CodeT5p model. We then use this embedding to retrieve the top-$k_s$ semantically similar snippets via HNSW from the pre-indexed repository vectors. Concurrently, the \textbf{structural path} performs a graph similarity search to find the top-$k_g$ subgraphs, that are structurally analogous to the query graph $G_q$. Finally, the candidate sets from both paths are merged and reranked based on a combined score to select the final top-$k$ most relevant graphs for the next phase.

\textbf{Phase 3: Graph Fusion Enhancement.}
This phase integrates the retrieved knowledge with the original query context. We extract dense node embeddings for the query graph $\mathbf{H}_q$ and each retrieved graph $\mathbf{H}_i$, then aggregate them into a context matrix $\mathbf{H}_r$ weighted by reranking scores. Next, we compute a cross-attention matrix $\mathbf{A}$ between query and retrieved nodes to identify relevant code parts. If the attention score between nodes exceeds a threshold and their types match, a cross-graph edge is added.

\textbf{Phase 4: Prompt Serialization.}
LLMs require linear text input, so we serialize the fused graph $G_f$ into a sequence of natural-language triples.

\textbf{Phase 5: LLM Inference.}
The current prompt includes both the original contextual information and the retrieved relevant context, which are provided to the LLM.

\begin{algorithm}[t]
   \caption{GRACE: Graph‐guided Repository-Aware Code Completion Pipeline}
   \label{alg:grace_pipeline}
   \begin{algorithmic}
   \REQUIRE
     Incomplete code snippet $\mathcal{C}$ with cursor position; full repository $\mathcal{R}$
   \ENSURE
     Predicted next line of code $\hat{\ell}$
   
   \STATE \textbf{// Phase 1: Query Graph Construction}
   \STATE Build AST of $\mathcal{C}$ and extract local symbols
   \STATE Construct query graph $G_q = (V_q,E_q)$ from AST nodes and edges
   
   \STATE \textbf{// Phase 2: Hybrid Graph Retrieval}
   \STATE Encode $G_q$ semantically (CodeT5p) $\rightarrow$ vector $\mathbf{v}_q$
   \STATE Encode repository graphs structurally (GNN + Laplacian) $\rightarrow$ index
   \STATE \textbf{Semantic path}: HNSW search $\rightarrow$ top-$k_s$ subgraphs $\mathcal{C}_{\text{sem}}$
   \STATE \textbf{Structural path}: Graph similarity search $\rightarrow$ top-$k_g$ subgraphs $\mathcal{C}_{\text{struct}}$
   \STATE Merge $\mathcal{C}_{\text{sem}} \cup \mathcal{C}_{\text{struct}}$ and rerank
   \STATE Select top-$k$ retrieved subgraphs $\{G_1,\dots,G_k\}$
   
   \STATE \textbf{// Phase 3: Graph Fusion Enhancement}
   \STATE Obtain node embeddings $\{\mathbf{H}_q,\mathbf{H}_1,\dots,\mathbf{H}_k\}$
   \STATE Aggregate retrieved node features $\mathbf{H}_r \leftarrow \sum_i w_i\mathbf{H}_i$
   \STATE Compute cross-attention $\mathbf{A} = \mathrm{softmax}(\mathbf{H}_q\mathbf{H}_r^\top/\sqrt{d})$
   \STATE Add cross-graph edges $(v_q,v_r)$ if $\mathbf{A}_{qr}>\theta$ and type-consistent
   \STATE Form fused graph $G_f =(V_q\cup V_r,E_q\cup E_r\cup E_{\text{cross}})$
   
   \STATE \textbf{// Phase 4: Prompt Serialization}
   \STATE Serialize $G_f$ to natural-language triples; concatenate with code context to obtain prompt $\mathcal{P}$
   
   \STATE \textbf{// Phase 5: LLM Inference}
   \STATE Query LLM with prompt $\mathcal{P}$; receive JSON output
   \STATE Parse field \texttt{completed\_code} as prediction $\hat{\ell}$
   
   \RETURN $\hat{\ell}$
   \end{algorithmic}
   \end{algorithm}

\section{Experiment}

\subsection{Experiment Setup}

\stitle{Datasets}: We employ two publicly available benchmarks that require cross-file reasoning:
\textbf{RepoEval-Updated}~\cite{liu2024graphcoder} and \textbf{CrossCodeEval}~\cite{ding2023crosscodeeval}.
\emph{CrossCodeEval} spans four languages (Python, Java, TypeScript, and~C); the Python split contains 2,665 test cases from 471 repositories, each demanding inter-file context to obtain the ground-truth line.
\emph{RepoEval-Updated} refreshes the original RepoEval\cite{zhang2023repocoder} by (i) removing repositories created before 2022-03-31 to mitigate training-data leakage and (ii) adding new Python and Java projects created up to 2023-01-01.
Tasks are categorised into \emph{line-level} and \emph{API-level} completion following~\cite{liu2023repobench}.

\stitle{LLMs}: We use Qwen2.5-Coder-14B, GPT-4o mini and DeepSeek-V3 as the backbone LLMs.

\stitle{Evaluation Metrics}: We adopt four complementary measures:
exact-match accuracy (\textbf{EM}), edit similarity (\textbf{ES}),
\textbf{Recall}, and the token-level \textbf{F1} score.
EM counts character-wise equality with the reference.
ES is $1-\text{Lev}/\max(|\text{pred}|,|\text{ref}|)$, where Lev denotes the Levenshtein distance.
Recall quantifies the proportion of reference tokens recovered, while F1 is the harmonic mean of precision and recall.

\stitle{Baselines}: We compare \modelname\ against five representative methods:\textbf{No RAG} (plain context completion),\textbf{Vanilla RAG} (sliding-window retrieval),
\textbf{GraphCoder}~\cite{liu2024graphcoder}, \textbf{RepoFuse}~\cite{liang2024repofuse}, and \textbf{RLCoder}~\cite{wang2024rlcoder}.
We try to adjust all baselines to the optimal parameters.

\stitle{Other Setup}: The total input window is fixed at 2,048 tokens, split evenly between retrieved context and local context.
RAG-based methods retrieve at most $k=10$ code snippets; the maximum generation length is 100 tokens.
\modelname\ encodes both textual and structural inputs via \textit{codet5p-110m-embedding} (768-dim), followed by HNSW (semantic path, $M{=}32$, $ef\_search{=}256$) and a flat Faiss index (structural path) for efficient nearest-neighbour search.
Attention threshold $\theta$ is set to~0.4 and the adaptive fusion weight $\alpha$ is tuned on a held-out validation split.
Experiments are conducted on NVIDIA A6000 GPU.

\stitle{RQs}: To comprehensively evaluate the performance of \modelname\, we propose the following research questions:

\begin{itemize}
  \item \textbf{RQ1-Effectiveness of GRACE:} Does \modelname\ outperform state-of-the-art baselines on repository-level code-completion benchmarks?

  \item \textbf{RQ2-Component Analysis:} How do individual architectural choices influence performance?  

  \item \textbf{RQ3-Model Scale Affection:} How does the backbone LLM size affect \modelname's effectiveness?

  \item \textbf{RQ4-Hyperparameter Sensitivity:} How sensitive is \modelname to the retrieval depth $k$?
\end{itemize}

\subsection{RQ1: Effectiveness of GRACE}

Table~\ref{tab:performance_comparison_combined} compares \modelname\ with six competitive baselines across two realistic benchmarks (\textsc{CrossCodeEval} and \textsc{RepoEval-Updated}), two programming languages (Python and Java), and three representative backbone LLMs (\textsc{GPT4o-mini}, \textsc{Qwen2.5-Coder-14B}, and \textsc{DeepSeek-V3}).  

\textbf{Overall superiority.}
Table~\ref{tab:performance_comparison_combined} shows that \modelname\ secures the top position in \textbf{35/48} metric–language–dataset combinations and ranks second in nine more, \textbf{consistently outperforming all six baselines}.  This demonstrates the robustness of graph-augmented retrieval and hierarchical fusion for practical repository-level completion.

\textbf{Effect of backbone size.}
Moving from a mid-size backbone (\textsc{Qwen2.5-Coder-14B}) to a larger one (\textsc{DeepSeek-V3}) widens the margin between \modelname\ and the best non-graph competitor from {\textbf{2.3\,pp}} to {\textbf{5.4\,pp}} (EM on Python/\textsc{CrossCodeEval}).  
This confirms our hypothesis that \emph{larger LLMs can better leverage the fine-grained structural cues surfaced by graph retrieval}.

\textbf{Dataset sensitivity.}
On \textsc{CrossCodeEval}, which stresses cross-file reasoning, \modelname\ achieves the highest \textit{F1} on \textbf{all 12\%} LLM–language pairs.  
On the more diverse \textsc{RepoEval-Updated}, it attains \textbf{83.95\% recall} in the Python subset—{\textbf{4.7\%}} higher than \textsc{Vanilla~RAG}—showcasing strong generalizability.

\textbf{Head-to-head with strong baselines.}
Although \textsc{RepoFuse} occasionally edges out \modelname\ in raw recall, its aggressive long-context retrieval inflates noise and lowers precision, depressing \textit{F1} by up to \textbf{8.6\%}.  
Under the largest backbone \textsc{DeepSeek-V3}, \modelname\ delivers an \textbf{average gain of +8.19\% EM} and \textbf{+7.51\% ES} over \textsc{RepoFuse}, and \textbf{+7.94\% EM} / \textbf{+6.83\% ES} over \textsc{RLCoder} across all datasets, highlighting its scalability advantages.

\textbf{Graph-structured retrieval coupled with hierarchical fusion boosts LLM code-completion accuracy, and the benefit amplifies with model scale and dataset complexity.}

\begin{table*}[htbp]
    \centering
    \caption{Performance comparison between \modelname\ and baselines on Different LLMs and Datasets}
    \label{tab:performance_comparison_combined}
    \resizebox{\textwidth}{!}{
    \setlength{\tabcolsep}{1pt}
    \begin{tabular}{@{}ll|cccccccc|cccccccc|cccccccc@{}}
    \toprule
     &  & \multicolumn{8}{c|}{GPT-4o mini} & \multicolumn{8}{c|}{Qwen2.5-Coder-14b} & \multicolumn{8}{c}{DeepSeek v3} \\
    \cmidrule(lr){3-10} \cmidrule(lr){11-18} \cmidrule(lr){19-26}
     &  & \multicolumn{4}{c}{Code} & \multicolumn{4}{c|}{Identifier} & \multicolumn{4}{c}{Code} & \multicolumn{4}{c|}{Identifier} & \multicolumn{4}{c}{Code} & \multicolumn{4}{c}{Identifier} \\
    \cmidrule(lr){3-6} \cmidrule(lr){7-10} \cmidrule(lr){11-14} \cmidrule(lr){15-18} \cmidrule(lr){19-22} \cmidrule(lr){23-26}
     & Method & EM & ES & Recall & F1 & EM & ES & Recall & F1 & EM & ES & Recall & F1 & EM & ES & Recall & F1 & EM & ES & Recall & F1 & EM & ES & Recall & F1 \\
    \midrule
    \multirow{6}{*}{\makecell[l]{CrossCode-\\Eval\\Python}} & No RAG & 5.59 & 55.40 & 77.79 & 75.92 & 10.47 & 55.85 & 44.56 & 42.73 & 1.16 & 35.89 & 83.64 & 68.44 & 1.84 & 38.48 & 40.77 & 27.77 & 9.53 & 44.48 & 88.49 & 73.09 & 14.07 & 47.22 & 53.52 & 37.46 \\
    & Vallina RAG & 9.31 & 58.54 & 78.74 & 77.39 & 15.95 & 58.97 & 48.80 & 47.28 & 2.85 & 40.53 & 83.68 & 70.19 & 4.47 & 42.62 & 45.11 & 32.96 & 13.32 & 47.31 & 88.05 & 73.95 & 18.87 & 50.58 & 53.95 & 40.80 \\
    & GraphCoder & 8.07 & 57.24 & 79.20 & 77.25 & 13.32 & 57.76 & 47.95 & 45.69 & 1.16 & 39.40 & 84.13 & 70.00 & 2.03 & 41.53 & 45.85 & 32.04 & 10.54 & 43.14 & 88.60 & 72.01 & 15.01 & 46.45 & 53.46 & 36.88 \\
    & RepoFuse & 22.36 & 66.12 & 83.52 & 81.67 & 30.96 & 67.86 & 62.41 & 59.84 & 5.33 & 45.88 & \underline{86.90} & 73.04 & 6.83 & 48.15 & 58.81 & 41.45 & 22.93 & 54.45 & 90.80 & 77.76 & 29.16 & 57.76 & 67.80 & 51.28 \\
    & RLCoder & \underline{25.44} & \underline{68.89} & \textbf{84.89} & \textbf{82.94} & \textbf{35.83} & \underline{70.97} & \underline{65.81} & \underline{63.37} & \underline{7.28} & \underline{47.95} & \textbf{87.26} & \underline{74.07} & \underline{9.87} & \underline{50.30} & \underline{58.89} & \underline{43.37} & \underline{26.45} & \underline{57.36} & \underline{90.62} & \underline{78.83} & \underline{34.30} & \underline{60.96} & \underline{68.05} & \underline{54.40} \\
    & Grace & \textbf{31.38} & \textbf{71.31} & \underline{83.69} & \underline{ 82.76} & \underline{32.85} & \textbf{73.64} & \textbf{74.71} & \textbf{74.63} & \textbf{13.96} & \textbf{63.32} & 80.46 & \textbf{77.67} & \textbf{15.38} & \textbf{61.31} & \textbf{63.69} & \textbf{72.76} & \textbf{32.85} & \textbf{63.64} & \textbf{94.71} & \textbf{83.96} & \textbf{43.96} & \textbf{73.32} & \textbf{80.46} & \textbf{67.67} \\
    \cmidrule{1-26}
    \multirow{6}{*}{\makecell[l]{CrossCode-\\Eval\\Java}} & No RAG & 12.01 & 63.52 & 79.01 & 79.55 & 19.82 & 62.80 & 52.66 & 52.69 & 9.21 & 59.64 & 79.42 & 77.81 & 16.78 & 60.62 & 52.37 & 50.42 & 21.18 & 68.22 & 84.61 & 83.29 & 29.27 & 68.33 & 62.11 & 59.74 \\
    & Vallina RAG & 15.85 & 65.57 & 79.75 & 80.32 & 24.17 & 64.77 & 55.50 & 55.76 & 11.64 & 61.54 & 81.11 & 79.26 & 20.29 & 62.66 & 54.80 & 52.71 & 25.99 & 70.23 & 84.91 & 83.91 & 35.06 & 70.44 & 64.41 & 62.72 \\
    & GraphCoder & 15.01 & 65.32 & 80.29 & 80.59 & 23.09 & 64.79 & 55.56 & 55.51 & 9.91 & 60.49 & 80.23 & 78.59 & 18.65 & 61.27 & 53.94 & 51.55 & 23.14 & 67.15 & 85.33 & 82.62 & 31.00 & 67.71 & 62.81 & 59.23 \\
    & RepoFuse & \underline{28.38} & 71.31 & 83.69 & \underline{83.76} & 39.08 & \underline{71.66} & \underline{65.36} & 65.15 & \underline{20.66} & \textbf{68.25} & \underline{83.93} & \textbf{82.53} & \textbf{33.61} & \underline{69.8} & 64.90 & 62.86 & 38.76 & 76.86 & 88.01 & 87.58 & 50.40 & 77.65 & 74.17 & 72.79 \\
    & RLCoder & \textbf{31.7} & \underline{72.72} & \underline{84.03} & \textbf{84.11} & \underline{41.51} & \textbf{72.67} & \textbf{66.93} & \underline{66.79} & \textbf{21.46} & \underline{68.07} & \textbf{84.20} & \underline{82.45} & 33.05 & 69.51 & \underline{65.05} & \underline{62.94} & \underline{40.91} & \underline{77.39} & \underline{88.38} & \underline{87.65} & \underline{51.99} & \underline{78.05} & \underline{74.38} & \underline{73.03} \\
    & Grace & 26.72 & \textbf{77.91} & \textbf{85.43} & 83.38 & \textbf{41.70} & 70.72 & 60.03 & \textbf{74.11} & 17.25 & 67.57 & 83.66 & 81.68 &\underline{32.85} & \textbf{73.64} & \textbf{74.71} & \textbf{75.03} & \textbf{42.37} & \textbf{79.59} & \textbf{90.82} & \textbf{90.15} & \textbf{55.19} & \textbf{79.15} & \textbf{85.03} & \textbf{89.21} \\
    \midrule[\heavyrulewidth]
    \midrule
    \multirow{6}{*}{\makecell[l]{RepoEval-\\Updated\\Python}} & No RAG & 21.35 & 50.07 & 74.53 & 69.51 & 29.50 & 53.30 & 54.11 & 46.92 & 22.20 & 50.06 & 76.23 & 70.29 & 29.65 & 53.53 & 54.68 & 46.91 & 29.45 & 59.41 & 84.07 & 77.57 & 37.90 & 62.43 & 66.75 & 57.21 \\
    & Vallina RAG & \underline{30.70} & 56.59 & 79.24 & 74.34 & 38.55 & 59.90 & 60.46 & 53.52 & \underline{29.60} & 54.09 & 80.26 & 73.67 & 36.55 & 57.47 & 59.14 & 51.18 & 37.45 & \underline{62.81} & \underline{86.86} & \underline{80.02} & 44.65 & \underline{66.10} & 70.55 & 61.15 \\
    & GraphCoder & 29.70 & \underline{58.86} & \underline{81.85} & \underline{76.49} & 38.70 & \underline{62.12} & \underline{64.42} & \underline{56.52} & 29.05 & \underline{56.93} & 80.54 & \underline{74.68} & 37.70 & \underline{60.54} & 63.16 & \underline{55.05} & 34.45 & 61.81 & 85.99 & 79.19 & 43.10 & 65.15 & 69.77 & 60.40 \\
    & RepoFuse & \textbf{32.10} & 56.25 & 80.00 & 74.19 & \textbf{39.65} & 59.47 & 62.79 & 54.45 & \textbf{32.80} & 55.81 & 80.69 & 74.08 & \textbf{40.25} & 58.90 & \underline{63.23} & 53.97 & \underline{38.85} & 62.44 & 86.25 & 79.20 & \underline{47.00} & 65.69 & \underline{72.69} & \underline{62.07} \\
    & RLCoder & 28.75 & 57.84 & 80.86 & 75.21 & 37.30 & 60.85 & 63.75 & 55.35 & 26.40 & 52.63 & \textbf{81.29} & 73.58 & 34.10 & 56.25 & 60.26 & 50.82 & 34.80 & 61.80 & 86.30 & 79.27 & 43.05 & 65.41 & 70.36 & 60.71 \\
    & Grace & 29.52 & \textbf{60.63} & \textbf{83.95} & \textbf{79.96} & \underline{39.15} & \textbf{68.54} & \textbf{66.52} & \textbf{60.23} & 29.20 & \textbf{57.45} & \underline{80.96} & \textbf{78.31} & \underline{38.47} & \textbf{62.83} & \textbf{68.69} & \textbf{70.08} & \textbf{57.26} & \textbf{76.25} & \textbf{89.22} & \textbf{85.20} & \textbf{52.70} & \textbf{70.43} & \textbf{77.59} & \textbf{72.39} \\
    \cmidrule{1-26}
    \multirow{6}{*}{\makecell[l]{RepoEval-\\Updated\\Java}} & No RAG & 17.70 & 56.25 & 75.29 & 75.22 & 26.50 & 57.08 & 55.80 & 51.08 & 17.80 & 55.95 & 76.29 & 75.30 & 26.95 & 57.39 & 57.30 & 51.84 & 26.10 & 63.85 & 80.45 & 79.59 & 36.15 & 64.62 & 65.85 & 60.02 \\
    & Vallina RAG & 23.45 & 58.81 & 76.43 & 76.19 & 31.40 & 59.25 & 58.80 & 53.94 & 21.80 & 56.43 & 76.75 & 75.43 & 29.65 & 57.37 & 57.56 & 51.88 & 30.60 & 63.77 & 80.98 & 79.62 & 39.50 & 64.80 & 66.68 & 60.35 \\
    & GraphCoder & \textbf{25.75} & \textbf{60.57} & 77.43 & 77.25 & \textbf{34.45} & \textbf{61.72} & \underline{61.16} & \underline{56.63} & \textbf{26.80} & \textbf{60.97} & 78.45 & \underline{77.87} & \underline{36.55} & \underline{62.21} & \underline{61.89} & \underline{57.08} & 30.80 & 64.38 & 81.15 & 79.93 & 39.50 & 65.32 & 66.77 & 61.09 \\
    & RepoFuse & 24.35 & 58.53 & 77.04 & 76.15 & 33.10 & 59.97 & 59.52 & 54.24 & \underline{24.10} & 58.65 & \underline{78.49} & 76.81 & 33.05 & 60.25 & 61.11 & 54.96 & \underline{32.55} & \underline{66.14} & \underline{81.70} & \underline{80.67} & \underline{43.10} & \underline{67.58} & \underline{69.30} & \underline{63.61} \\
    & RLCoder & \underline{25.30} & \underline{60.40} & \underline{77.80} & \underline{77.39} & \underline{34.30} & \underline{61.41} & 60.89 & 56.13 & 23.60 & 57.98 & 78.02 & 76.48 & 31.80 & 59.13 & 59.78 & 54.08 & 31.95 & 66.08 & 81.28 & 80.58 & 41.20 & 66.59 & 68.51 & 63.25 \\
    & Grace & 23.85 & 59.38 & \textbf{81.15} & \textbf{79.93} & 32.50 & 60.32 & \textbf{66.77} & \textbf{61.09} & 23.2 & \underline{59.85} & \textbf{80.96} & \textbf{78.31} & \textbf{38.47} & \textbf{62.83} & \textbf{68.69} & \textbf{70.08} & \textbf{33.37} & \textbf{70.46} & \textbf{84.62} & \textbf{86.27} & \textbf{45.21} & \textbf{67.37} & \textbf{72.36} & \textbf{70.52} \\
   \bottomrule
   \end{tabular}%
   }
   \end{table*}

\subsection{RQ2: Component Analysis}\label{sec:rq2}

   \begin{table}[ht]
   \centering
   \caption{Ablation study of \modelname.  BM25: replace hybrid graph retriever with BM25; w/o~Fusion: remove graph–fusion stage and simply concatenate retrieved code; AST-Only: build graphs from abstract syntax trees only.  Metrics are macro-averaged over \textsc{Code} and \textsc{Identifier} tasks.}
   \label{tab:ablation}
   \setlength{\tabcolsep}{1pt}
   \begin{tabular}{@{}llcccc@{}}
   \toprule
   \textbf{Dataset} & \textbf{Metric} & \textbf{\modelname} & \textbf{BM25} & \textbf{w/o Fusion} & \textbf{AST-Only} \\
   \midrule
   \multirow{2}{*}{\makecell[l]{CrossCodeEval\\Python}}   & EM & \textbf{31.38} & {27.36} & {25.98} & \underline{28.22} \\
                                       & F1 & \textbf{82.76} & {78.97} & 76.59 & \underline{79.48} \\ \midrule
   \multirow{2}{*}{\makecell[l]{CrossCodeEval\\Java}} & EM & \textbf{26.72} & {23.59} & 22.36 & \underline{24.75} \\
                                       & F1 & \textbf{83.38} & {80.33} & 79.11 & \underline{81.24} \\ \midrule
   \multirow{2}{*}{\makecell[l]{RepoEval-\\Updated Python}}    & EM & \textbf{29.52} & {26.48} & 25.7 & \underline{27.52} \\
                                       & F1 & \textbf{79.96} & {75.82} & 74.12 & \underline{76.75} \\ \midrule
   \multirow{2}{*}{\makecell[l]{RepoEval-\\Updated Java}}  & EM & \textbf{23.85} & \underline{21.40} & 20.30 & {21.19} \\
                                       & F1 & \textbf{79.93} & {77.13} & 76.24 & \underline{78. 89} \\
   \bottomrule
   \end{tabular}
   \end{table}
   
   Table~\ref{tab:ablation} reveals the contribution of each design choice in GRACE.  

   \noindent\textbf{Key findings are as follows:}
   
(1) \emph{Hybrid Graph Retriever matters.} Substituting it with a lexical BM25 search cuts \textbf{F1 by 4–6\%} (e.g.\ \textbf{82.76\%\,→\,78.97\%} on CrossCodeEval-Py), showing that purely textual retrieval cannot surface the structurally relevant context.

(2) \emph{Graph fusion is decisive.} Removing the fusion stage (“w/o Fusion’’) causes the sharpest drop—up to \textbf{–6.2\% F1} and \textbf{–5.4\% EM}—because naïve concatenation fails to reconcile overlaps and noise among retrieved snippets.

(3) \emph{Rich relations beyond AST are still needed.} Limiting the graph to AST edges closes part of the gap but remains \textbf{2–3\%} behind full GRACE, confirming that call-, data- and control-flow links offer complementary cues.

Together, these ablations underline three indispensable pillars of GRACE: precise hybrid retrieval, hierarchical graph fusion, and a multi-relation code graph—all required to reach the \textbf{state-of-the-art 80–83\% F1} achieved by the complete system.

\subsection{RQ3: Model Scale Affection}
\label{sec:rq3}

\begin{figure}[htbp]
\centering
\includegraphics[width=\linewidth]{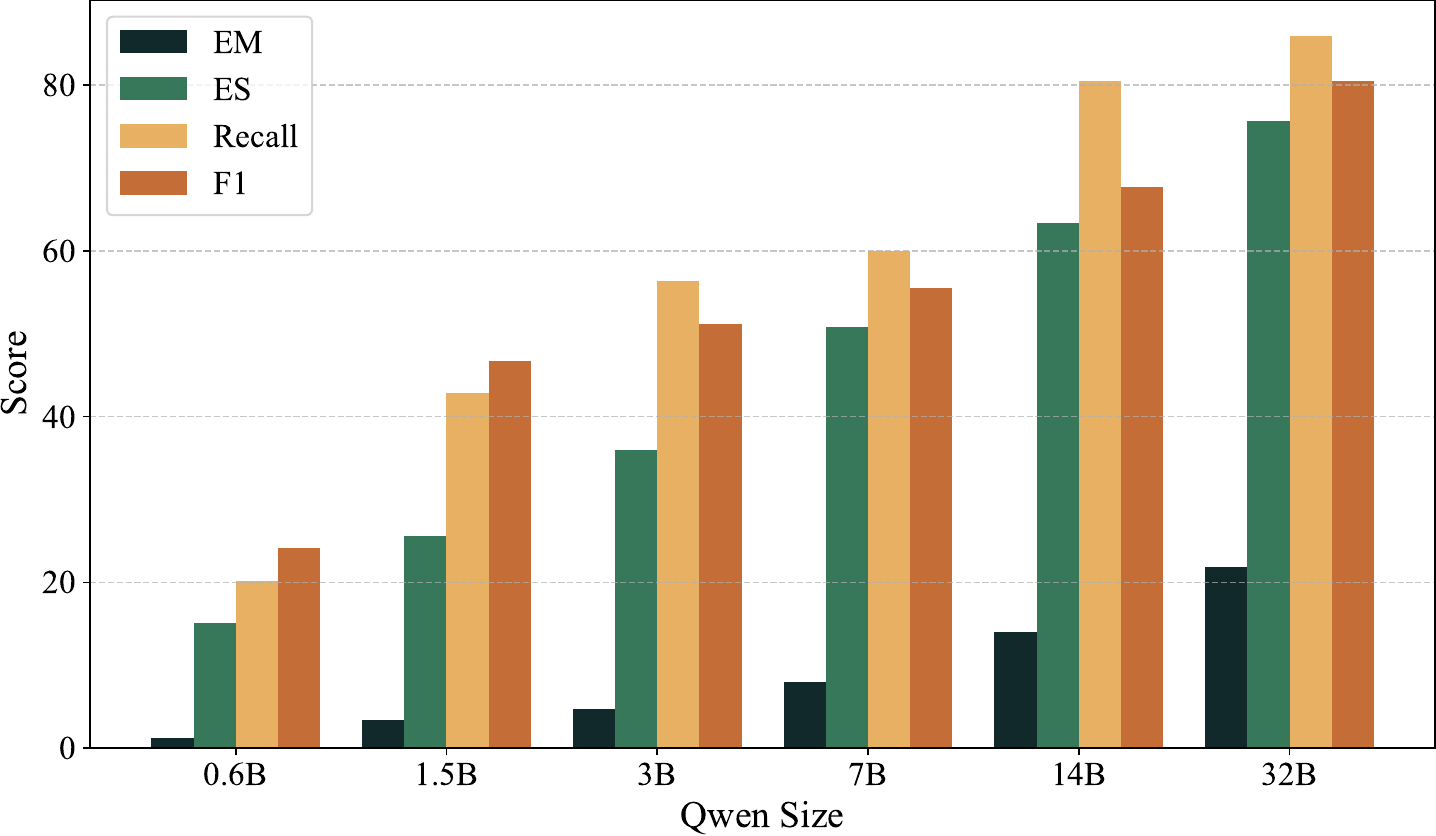}
\caption{Impact of Qwen backbone size on the performance of \modelname on CrossCodeEval dataset.}
\label{fig:modelsize}
\end{figure}

We examine how the capacity of the backbone LLM influences the effectiveness of \modelname.  
Figure~\ref{fig:modelsize} reports results on six Qwen variants ranging from 0.6B to 32B parameters.  
\noindent\textbf{Key findings.} Scaling the Qwen backbone from \textbf{0.6B} to \textbf{32B} parameters boosts \textit{F1} by \textbf{+56.3\%} (24.20\% to \textbf{80.46\%}), \textit{EM} by \textbf{+15.1\%}, and \textit{ES} by a remarkable \textbf{+75.7\%}.
A closer inspection shows two distinct phases.
\begin{itemize}
    \item \emph{Sub-billion to 7B.} Gains are steady but moderate: every doubling of model size yields roughly +10\% in ES.
    \item \emph{14B → 32B.} Improvements accelerate; ES jumps by +12.4\% and F1 by +12.8\%, indicating that larger models extract and leverage the graph-structured cues far more effectively.
\end{itemize}

Together with RQ1, this confirms that \modelname\ not only scales gracefully but \emph{synergises} with high-capacity LLMs, pointing to an orthogonal avenue of progress relative to pure parameter scaling.

\subsection{RQ4: Hyperparameter Sensitivity}

\label{sec:rq4}

\begin{figure}[htbp]
\centering
\includegraphics[width=\linewidth]{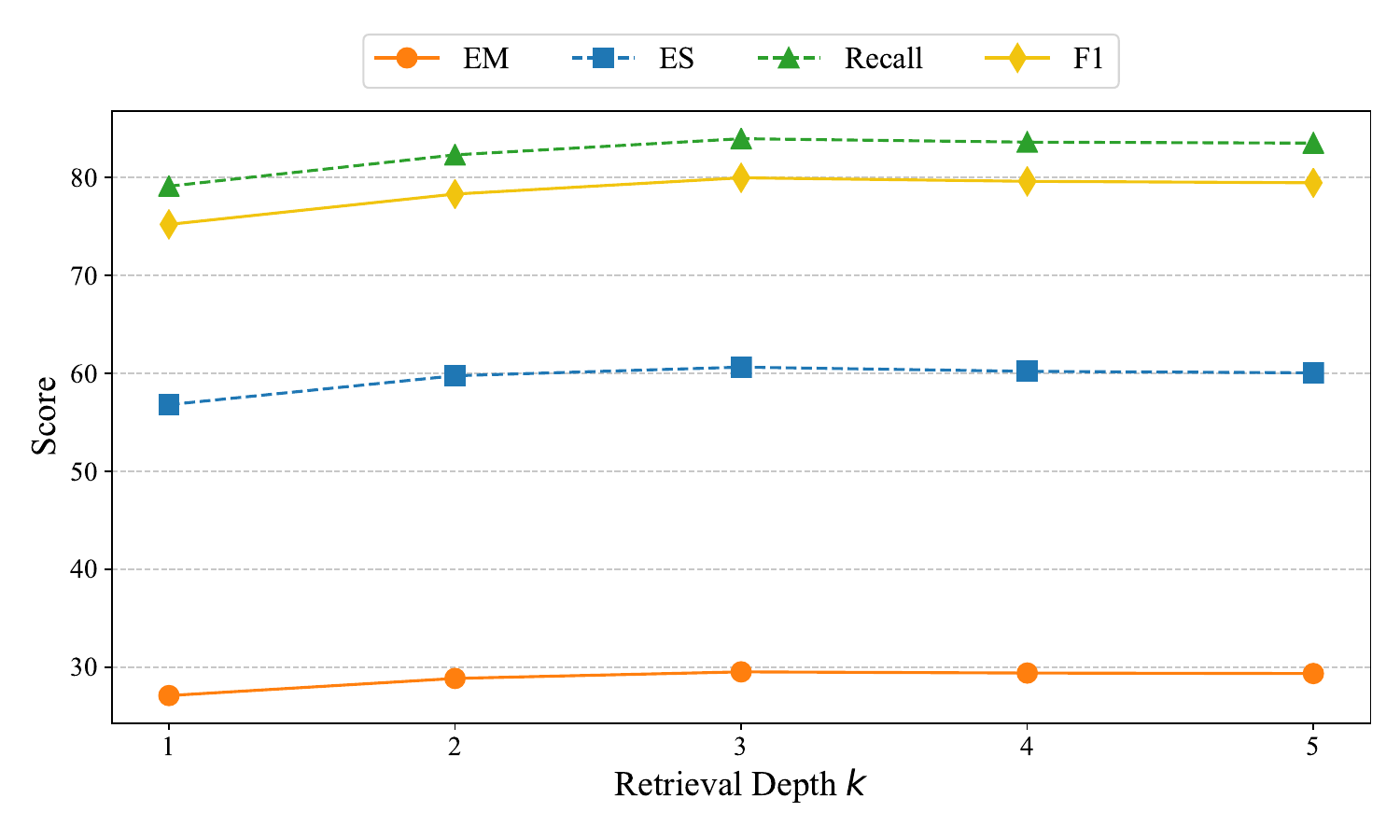}
\caption{Effect of retrieval depth $k$ on the macro performance of \modelname.}
\label{fig:topk}
\end{figure}

We study how the retrieval depth $k$ affects \modelname's performance on RepoEval-Updated dataset.  
As shown in Figure~\ref{fig:topk}, enlarging $k$ from 1 to 3 consistently boosts all four metrics, because a broader candidate set increases the likelihood of retrieving truly relevant context.  
Beyond $k=3$, the curves flatten: \textit{F1} varies within $0.5\%$ and \textit{EM} within $0.2\%$ between $k=3$ and $k=5$.  
This plateau indicates that our hierarchical fusion module is already able to distill sufficient information from the top-3 subgraphs, and that adding more neighbours mostly introduces redundant edges rather than novel cues.  
Therefore, we adopt $k=3$ as the default, striking a good balance between accuracy and retrieval cost.

\section{Discussion}

\subsection{Threats to Validity}

\stitle{Internal validity:} (1) \textit{Implementation correctness}. \modelname comprises several non-trivial components (graph extractor, hybrid retriever, fusion module). Implementation bugs or sub-optimal engineering choices could artificially inflate or deflate performance. We have released the full source code to facilitate auditing and reproduction.

\stitle{External validity:} (1) \textit{Dataset representativeness}. Our experiments rely on two public repository-level completion benchmarks drawn mainly from Python and Java projects. Results may not generalise to other languages (e.g., C/C++,  Go, Rust) or industrial monorepos with proprietary frameworks. Extending \modelname to additional ecosystems forms part of future work.
(2) \textit{Repository scale}. Although the chosen corpora contain hundreds of files, very large codebases (millions of LOC) could stress graph construction time or memory. Our complexity analysis highlights potential bottlenecks; we plan incremental indexing strategies to mitigate this threat.

\stitle{Construct validity:} (1) \textit{Evaluation metrics}. We adopt standard EM, ES, F1 and Recall, which capture syntactic correctness but not semantic soundness (e.g., compilation success). Complementary metrics such as pass@k on unit tests or human assessment could yield different conclusions.
(2) \textit{Baseline selection}. We compare against strong text-centric RAG baselines but omit approaches that leverage IDE signals or test-aware feedback. While orthogonal to our contribution, their inclusion might narrow the observed performance gap.

\subsection{Complexity Analysis}

This section analyzes the complexity of the proposed graph fusion enhancement mechanism. Let $n_q=|V_q|$ and $|E_q|$ denote the numbers of nodes and edges in the query graph, and $n_r=|V_r|=\sum_{i=1}^{k} |V_i|$, $|E_r|=\sum_{i=1}^{k}|E_i|$ those of the $k$ retrieved sub-graphs.  $L$ is the number of GNN layers and $d$ the hidden dimension.

\textbf{Phase 1: Query graph construction.}  
Parsing the incomplete snippet into an AST and building local CFG/DFG is linear in the source length, denoted $O(|\mathcal{C}|)$.  As $|\mathcal{C}| \!\approx\! n_q$, we treat this as $O(n_q)$.

\textbf{Phase 2: Hybrid graph retrieval.}  
Semantic encoding of $G_q$ uses a GNN: $O(L\,|E_q|\,d)$.  
Semantic ANN search on an HNSW index over $N_s$ snippets costs $O(\log N_s + k_s)$.  
Structural encoding of $G_q$ (concatenating node embedding and Laplacian PE) is $O(L\,|E_q|\,d)$;  
graph-vector search over $N_g$ indexed sub-graphs is $O(\log N_g + k_g)$.  
Overall retrieval complexity:
\[
O\!\bigl(L\,|E_q|\,d + \log N_s + \log N_g + k_s + k_g \bigr).
\]

\textbf{Phase 3: Graph fusion enhancement.}  
Computing node embeddings for the $k$ retrieved graphs: $O\!\bigl(L\,|E_r|\,d\bigr)$.  
Cross-attention between query and retrieved nodes dominates: $O(n_q\,n_r\,d)$.  
Edge construction and filtering add $O(n_q\,n_r)$.

\textbf{Phase 4: Prompt serialization.}  
Serialising the fused graph is linear in its size: $O(n_q + n_r + |E_q| + |E_r|)$.

\textbf{Phase 5: LLM inference.}  
This depends on prompt length $T$ (tokens) and LLM architecture; we denote it $O(\mathrm{LLM}(T))$.

\textbf{Overall time complexity.}  
The end-to-end pipeline is therefore bounded by
\[
O\!\bigl(L\,(|E_q| + |E_r|)\,d \;+\; n_q\,n_r\,d \;+\; \log N_s + \log N_g + k_s + k_g \;+\; \mathrm{LLM}(T)\bigr).
\]
In practice, the cross-attention term $O(n_q\,n_r\,d)$ becomes the primary bottleneck when the retrieved context is large, while HNSW search and GNN encoding remain sub-linear or linear in repository size due to indexing.
\section{Related Work}
\subsection{Large Language Models for Code}
In recent years, the rapid advancement of Large Language Models (LLMs) has significantly enhanced the performance of software engineering tasks such as code completion\cite{achiam2023gpt,guo2024deepseek,luo2023wizardcoder,wang2023codet5+} . These models are broadly classified into closed-source and open-source variants.

Proprietary models, such as OpenAI's GPT-4\cite{openai2024gpt4o}, which was developed upon its pioneering model Codex\cite{chen2021evaluating}, along with Google's Gemini 2.5\cite{kavukcuoglu2025gemini} and Anthropic's Claude 3.5 Sonnet\cite{anthropic2024claude}, have achieved leading positions on multiple code benchmarks like HumanEval\cite{chen2021evaluating} and MBPP\cite{austin2021program}. Concurrently, the open-source community has witnessed the emergence of numerous high-performance models. This includes powerful general-purpose LLMs with strong coding capabilities, such as Llama 3\cite{grattafiori2024llama}, Qwen 2.5\cite{qwen2025qwen25technicalreport}, and DeepSeek-V3\cite{liu2024deepseek}, as well as dedicated code models like Code Llama\cite{roziere2023code}, StarCoder\cite{li2023starcoder,lozhkov2024starcoder}, and DeepSeek-Coder\cite{guo2024deepseek,zhu2024deepseek}.The prevailing paradigm for these models is to represent source code as a linear sequence of tokens, with training centered on autoregressive objectives like Next Token Prediction or Fill-in-the-Middle\cite{bavarian2022efficient}.

Despite their proficiency in generating syntactically correct code snippets, these models perform suboptimally on repository-level code completion tasks, which better reflect real-world software development scenarios, due to their limited context windows\cite{liu2023lost}.

\subsection{Repository-level Code Completion}
In real-world software development, coding often requires referencing the contextual information of the entire codebase. Consequently, the task of repository-level code completion, which effectively leverages this information, is gaining increasing attention from both academia and industry\cite{liu2023repobench,ding2023crosscodeeval}. Although LLMs have demonstrated powerful capabilities in code generation, their inherent context window limitations make it difficult for them to directly process large-scale codebases\cite{zhang2023repocoder,ding2022cocomic}.

To overcome this limitation, researchers have introduced the Retrieval-Augmented Generation (RAG) framework\cite{lewis2020retrieval,khandelwal2019generalization,li2023acecoder,zan2022language,zhang2023syntax,tan2024prompt}. AceCoder\cite{li2023acecoder} finds relevant reference examples by searching the codebase for programs similar to the current requirement, helping the model generate more accurate code. APICoder\cite{zan2022language}, which includes an APIRetriever and an APICoder, retrieves potentially useful API information from a private library's API documentation based on the task; this information is then passed to the APICoder as a reference for code generation. kNN-TRANX\cite{zhang2023syntax} is a token-level retrieval-augmented code generation method that improves code generation performance while reducing retrieval noise. RepoCoder\cite{zhang2023repocoder} proposes an iterative retrieval-generation pipeline to further enhance the performance of standard RAG methods. While RAG effectively extends the model's contextual scope, these methods primarily rely on semantic similarity for retrieval and often overlook the structural information of the code.

Consequently, a growing body of research has attempted to represent code as a graph structure to capture its complex dependencies. For example, GraphCoder\cite{liu2024graphcoder} captures the contextual information of a codebase through a Code Context Graph (CCG) and then employs a two-stage coarse-to-fine retrieval strategy. RepoHyper\cite{phan2024repohyper} constructs a repository-level semantic graph and performs graph expansion and retrieval operations upon it to obtain more precise structured context. CoCoMIC\cite{ding2022cocomic} utilizes a Dependency Graph search from the field of program analysis; during completion, it first locates the node corresponding to the segment to be completed and then treats its neighboring nodes as supplementary context. RepoFuse\cite{liang2024repofuse} proposes a Repo-specific Semantic Graph, retrieves both Semantic Context and Similar Context, and filters for the most beneficial context through a Relevance-Guided Context Selection method\cite{phan2024repohyper,cheng2024dataflow}.

Furthermore, another line of work focuses on how to integrate the retrieved information into the generation model most efficiently. For instance, methods like RepoFusion\cite{shrivastava2023repofusion} and RepoPrompts\cite{shrivastava2023repository} have proposed different strategies to optimize the interaction between the retrieved information and the LLM, aiming to maximize its assistive effect on the generation process.
\section{Conclusion}


This paper presents \modelname, an innovative graph-based retrieval-augmented code completion framework that systematically addresses the critical limitations of existing repository-level code completion techniques. By modeling code repositories as hierarchical, semantically-rich code graph databases, we fundamentally transform the limited perspective of traditional RAG methods. The core innovations of the GRACE framework are manifested in three aspects: First, we construct a multi-level, multi-semantic hybrid code graph that comprehensively captures structural information at different granularities from repository-level to function-level; Second, our designed hybrid graph retriever significantly improves retrieval relevance by integrating GNN-based structural retrieval with text semantic retrieval; Finally, our proposed graph fusion enhancement mechanism achieves, for the first time, effective utilization of structural associations between retrieved code and code-to-be-completed. Experimental results validate the effectiveness of \modelname. Compared to existing state-of-the-art methods, \modelname demonstrate the critical value of graph structural information for repository-level code completion tasks. Future work will explore dynamic graph updating for evolving codebases and generalize \modelname to multilingual software environments.


   
   
    
    
    
    
    


\clearpage
\bibliographystyle{ACM-Reference-Format}
\bibliography{references}


\begin{thebibliography}{58}


\ifx \showCODEN    \undefined \def \showCODEN     #1{\unskip}     \fi
\ifx \showISBNx    \undefined \def \showISBNx     #1{\unskip}     \fi
\ifx \showISBNxiii \undefined \def \showISBNxiii  #1{\unskip}     \fi
\ifx \showISSN     \undefined \def \showISSN      #1{\unskip}     \fi
\ifx \showLCCN     \undefined \def \showLCCN      #1{\unskip}     \fi
\ifx \shownote     \undefined \def \shownote      #1{#1}          \fi
\ifx \showarticletitle \undefined \def \showarticletitle #1{#1}   \fi
\ifx \showURL      \undefined \def \showURL       {\relax}        \fi
\providecommand\bibfield[2]{#2}
\providecommand\bibinfo[2]{#2}
\providecommand\natexlab[1]{#1}
\providecommand\showeprint[2][]{arXiv:#2}

\bibitem[Achiam et~al\mbox{.}(2023)]%
        {achiam2023gpt}
\bibfield{author}{\bibinfo{person}{Josh Achiam}, \bibinfo{person}{Steven Adler}, \bibinfo{person}{Sandhini Agarwal}, \bibinfo{person}{Lama Ahmad}, \bibinfo{person}{Ilge Akkaya}, \bibinfo{person}{Florencia~Leoni Aleman}, \bibinfo{person}{Diogo Almeida}, \bibinfo{person}{Janko Altenschmidt}, \bibinfo{person}{Sam Altman}, \bibinfo{person}{Shyamal Anadkat}, {et~al\mbox{.}}} \bibinfo{year}{2023}\natexlab{}.
\newblock \showarticletitle{Gpt-4 technical report}.
\newblock \bibinfo{journal}{\emph{arXiv preprint arXiv:2303.08774}} (\bibinfo{year}{2023}).
\newblock


\bibitem[Agrawal et~al\mbox{.}(2023)]%
        {agrawal2023guiding}
\bibfield{author}{\bibinfo{person}{Lakshya~A Agrawal}, \bibinfo{person}{Aditya Kanade}, \bibinfo{person}{Navin Goyal}, \bibinfo{person}{Shuvendu~K Lahiri}, {and} \bibinfo{person}{Sriram~K Rajamani}.} \bibinfo{year}{2023}\natexlab{}.
\newblock \showarticletitle{Guiding language models of code with global context using monitors}.
\newblock \bibinfo{journal}{\emph{arXiv preprint arXiv:2306.10763}} (\bibinfo{year}{2023}).
\newblock


\bibitem[Allal et~al\mbox{.}(2023)]%
        {allal2023santacoder}
\bibfield{author}{\bibinfo{person}{Loubna~Ben Allal}, \bibinfo{person}{Raymond Li}, \bibinfo{person}{Denis Kocetkov}, \bibinfo{person}{Chenghao Mou}, \bibinfo{person}{Christopher Akiki}, \bibinfo{person}{Carlos~Munoz Ferrandis}, \bibinfo{person}{Niklas Muennighoff}, \bibinfo{person}{Mayank Mishra}, \bibinfo{person}{Alex Gu}, \bibinfo{person}{Manan Dey}, {et~al\mbox{.}}} \bibinfo{year}{2023}\natexlab{}.
\newblock \showarticletitle{Santacoder: don't reach for the stars!}
\newblock \bibinfo{journal}{\emph{arXiv preprint arXiv:2301.03988}} (\bibinfo{year}{2023}).
\newblock


\bibitem[Anthropic(2024)]%
        {anthropic2024claude}
\bibfield{author}{\bibinfo{person}{Anthropic}.} \bibinfo{year}{2024}\natexlab{}.
\newblock \bibinfo{title}{Claude 3.5 Sonnet}.
\newblock \bibinfo{howpublished}{\url{https://www.anthropic.com/news/claude-3-5-sonnet}}.
\newblock
\newblock
\shownote{2024}.


\bibitem[Austin et~al\mbox{.}(2021)]%
        {austin2021program}
\bibfield{author}{\bibinfo{person}{Jacob Austin}, \bibinfo{person}{Augustus Odena}, \bibinfo{person}{Maxwell Nye}, \bibinfo{person}{Maarten Bosma}, \bibinfo{person}{Henryk Michalewski}, \bibinfo{person}{David Dohan}, \bibinfo{person}{Ellen Jiang}, \bibinfo{person}{Carrie Cai}, \bibinfo{person}{Michael Terry}, \bibinfo{person}{Quoc Le}, {et~al\mbox{.}}} \bibinfo{year}{2021}\natexlab{}.
\newblock \showarticletitle{Program synthesis with large language models}.
\newblock \bibinfo{journal}{\emph{arXiv preprint arXiv:2108.07732}} (\bibinfo{year}{2021}).
\newblock


\bibitem[Bairi et~al\mbox{.}(2024)]%
        {bairi2024codeplan}
\bibfield{author}{\bibinfo{person}{Ramakrishna Bairi}, \bibinfo{person}{Atharv Sonwane}, \bibinfo{person}{Aditya Kanade}, \bibinfo{person}{Vageesh~D C}, \bibinfo{person}{Arun Iyer}, \bibinfo{person}{Suresh Parthasarathy}, \bibinfo{person}{Sriram Rajamani}, \bibinfo{person}{Balasubramanyan Ashok}, {and} \bibinfo{person}{Shashank Shet}.} \bibinfo{year}{2024}\natexlab{}.
\newblock \showarticletitle{Codeplan: Repository-level coding using llms and planning}.
\newblock \bibinfo{journal}{\emph{Proceedings of the ACM on Software Engineering}} \bibinfo{volume}{1}, \bibinfo{number}{FSE} (\bibinfo{year}{2024}), \bibinfo{pages}{675--698}.
\newblock


\bibitem[Bavarian et~al\mbox{.}(2022)]%
        {bavarian2022efficient}
\bibfield{author}{\bibinfo{person}{Mohammad Bavarian}, \bibinfo{person}{Heewoo Jun}, \bibinfo{person}{Nikolas Tezak}, \bibinfo{person}{John Schulman}, \bibinfo{person}{Christine McLeavey}, \bibinfo{person}{Jerry Tworek}, {and} \bibinfo{person}{Mark Chen}.} \bibinfo{year}{2022}\natexlab{}.
\newblock \showarticletitle{Efficient training of language models to fill in the middle}.
\newblock \bibinfo{journal}{\emph{arXiv preprint arXiv:2207.14255}} (\bibinfo{year}{2022}).
\newblock


\bibitem[Chen et~al\mbox{.}(2021)]%
        {chen2021evaluating}
\bibfield{author}{\bibinfo{person}{Mark Chen}, \bibinfo{person}{Jerry Tworek}, \bibinfo{person}{Heewoo Jun}, \bibinfo{person}{Qiming Yuan}, \bibinfo{person}{Henrique Ponde De~Oliveira Pinto}, \bibinfo{person}{Jared Kaplan}, \bibinfo{person}{Harri Edwards}, \bibinfo{person}{Yuri Burda}, \bibinfo{person}{Nicholas Joseph}, \bibinfo{person}{Greg Brockman}, {et~al\mbox{.}}} \bibinfo{year}{2021}\natexlab{}.
\newblock \showarticletitle{Evaluating large language models trained on code}.
\newblock \bibinfo{journal}{\emph{arXiv preprint arXiv:2107.03374}} (\bibinfo{year}{2021}).
\newblock


\bibitem[Cheng et~al\mbox{.}(2024)]%
        {cheng2024dataflow}
\bibfield{author}{\bibinfo{person}{Wei Cheng}, \bibinfo{person}{Yuhan Wu}, {and} \bibinfo{person}{Wei Hu}.} \bibinfo{year}{2024}\natexlab{}.
\newblock \showarticletitle{Dataflow-guided retrieval augmentation for repository-level code completion}.
\newblock \bibinfo{journal}{\emph{arXiv preprint arXiv:2405.19782}} (\bibinfo{year}{2024}).
\newblock


\bibitem[Chowdhery et~al\mbox{.}(2023)]%
        {chowdhery2023palm}
\bibfield{author}{\bibinfo{person}{Aakanksha Chowdhery}, \bibinfo{person}{Sharan Narang}, \bibinfo{person}{Jacob Devlin}, \bibinfo{person}{Maarten Bosma}, \bibinfo{person}{Gaurav Mishra}, \bibinfo{person}{Adam Roberts}, \bibinfo{person}{Paul Barham}, \bibinfo{person}{Hyung~Won Chung}, \bibinfo{person}{Charles Sutton}, \bibinfo{person}{Sebastian Gehrmann}, {et~al\mbox{.}}} \bibinfo{year}{2023}\natexlab{}.
\newblock \showarticletitle{Palm: Scaling language modeling with pathways}.
\newblock \bibinfo{journal}{\emph{Journal of Machine Learning Research}} \bibinfo{volume}{24}, \bibinfo{number}{240} (\bibinfo{year}{2023}), \bibinfo{pages}{1--113}.
\newblock


\bibitem[Christopoulou et~al\mbox{.}(2022)]%
        {christopoulou2022pangu}
\bibfield{author}{\bibinfo{person}{Fenia Christopoulou}, \bibinfo{person}{Gerasimos Lampouras}, \bibinfo{person}{Milan Gritta}, \bibinfo{person}{Guchun Zhang}, \bibinfo{person}{Yinpeng Guo}, \bibinfo{person}{Zhongqi Li}, \bibinfo{person}{Qi Zhang}, \bibinfo{person}{Meng Xiao}, \bibinfo{person}{Bo Shen}, \bibinfo{person}{Lin Li}, {et~al\mbox{.}}} \bibinfo{year}{2022}\natexlab{}.
\newblock \showarticletitle{Pangu-coder: Program synthesis with function-level language modeling}.
\newblock \bibinfo{journal}{\emph{arXiv preprint arXiv:2207.11280}} (\bibinfo{year}{2022}).
\newblock


\bibitem[Clement et~al\mbox{.}(2021)]%
        {clement2021long}
\bibfield{author}{\bibinfo{person}{Colin~B Clement}, \bibinfo{person}{Shuai Lu}, \bibinfo{person}{Xiaoyu Liu}, \bibinfo{person}{Michele Tufano}, \bibinfo{person}{Dawn Drain}, \bibinfo{person}{Nan Duan}, \bibinfo{person}{Neel Sundaresan}, {and} \bibinfo{person}{Alexey Svyatkovskiy}.} \bibinfo{year}{2021}\natexlab{}.
\newblock \showarticletitle{Long-range modeling of source code files with eWASH: Extended window access by syntax hierarchy}.
\newblock \bibinfo{journal}{\emph{arXiv preprint arXiv:2109.08780}} (\bibinfo{year}{2021}).
\newblock


\bibitem[Ding et~al\mbox{.}(2023)]%
        {ding2023crosscodeeval}
\bibfield{author}{\bibinfo{person}{Yangruibo Ding}, \bibinfo{person}{Zijian Wang}, \bibinfo{person}{Wasi Ahmad}, \bibinfo{person}{Hantian Ding}, \bibinfo{person}{Ming Tan}, \bibinfo{person}{Nihal Jain}, \bibinfo{person}{Murali~Krishna Ramanathan}, \bibinfo{person}{Ramesh Nallapati}, \bibinfo{person}{Parminder Bhatia}, \bibinfo{person}{Dan Roth}, {et~al\mbox{.}}} \bibinfo{year}{2023}\natexlab{}.
\newblock \showarticletitle{Crosscodeeval: A diverse and multilingual benchmark for cross-file code completion}.
\newblock \bibinfo{journal}{\emph{Advances in Neural Information Processing Systems}}  \bibinfo{volume}{36} (\bibinfo{year}{2023}), \bibinfo{pages}{46701--46723}.
\newblock


\bibitem[Ding et~al\mbox{.}(2022)]%
        {ding2022cocomic}
\bibfield{author}{\bibinfo{person}{Yangruibo Ding}, \bibinfo{person}{Zijian Wang}, \bibinfo{person}{Wasi~Uddin Ahmad}, \bibinfo{person}{Murali~Krishna Ramanathan}, \bibinfo{person}{Ramesh Nallapati}, \bibinfo{person}{Parminder Bhatia}, \bibinfo{person}{Dan Roth}, {and} \bibinfo{person}{Bing Xiang}.} \bibinfo{year}{2022}\natexlab{}.
\newblock \showarticletitle{Cocomic: Code completion by jointly modeling in-file and cross-file context}.
\newblock \bibinfo{journal}{\emph{arXiv preprint arXiv:2212.10007}} (\bibinfo{year}{2022}).
\newblock


\bibitem[Grattafiori et~al\mbox{.}(2024)]%
        {grattafiori2024llama}
\bibfield{author}{\bibinfo{person}{Aaron Grattafiori}, \bibinfo{person}{Abhimanyu Dubey}, \bibinfo{person}{Abhinav Jauhri}, \bibinfo{person}{Abhinav Pandey}, \bibinfo{person}{Abhishek Kadian}, \bibinfo{person}{Ahmad Al-Dahle}, \bibinfo{person}{Aiesha Letman}, \bibinfo{person}{Akhil Mathur}, \bibinfo{person}{Alan Schelten}, \bibinfo{person}{Alex Vaughan}, {et~al\mbox{.}}} \bibinfo{year}{2024}\natexlab{}.
\newblock \showarticletitle{The llama 3 herd of models}.
\newblock \bibinfo{journal}{\emph{arXiv preprint arXiv:2407.21783}} (\bibinfo{year}{2024}).
\newblock


\bibitem[Guo et~al\mbox{.}(2024)]%
        {guo2024deepseek}
\bibfield{author}{\bibinfo{person}{Daya Guo}, \bibinfo{person}{Qihao Zhu}, \bibinfo{person}{Dejian Yang}, \bibinfo{person}{Zhenda Xie}, \bibinfo{person}{Kai Dong}, \bibinfo{person}{Wentao Zhang}, \bibinfo{person}{Guanting Chen}, \bibinfo{person}{Xiao Bi}, \bibinfo{person}{Yu Wu}, \bibinfo{person}{YK Li}, {et~al\mbox{.}}} \bibinfo{year}{2024}\natexlab{}.
\newblock \showarticletitle{DeepSeek-Coder: When the Large Language Model Meets Programming--The Rise of Code Intelligence}.
\newblock \bibinfo{journal}{\emph{arXiv preprint arXiv:2401.14196}} (\bibinfo{year}{2024}).
\newblock


\bibitem[Kavukcuoglu(2025)]%
        {kavukcuoglu2025gemini}
\bibfield{author}{\bibinfo{person}{Koray Kavukcuoglu}.} \bibinfo{year}{2025}\natexlab{}.
\newblock \bibinfo{title}{Gemini 2.5: Our most intelligent AI model}.
\newblock \bibinfo{howpublished}{\url{https://blog.google/technology/google-deepmind/gemini-model-thinking-updates-march-2025/}}.
\newblock
\newblock
\shownote{2025}.


\bibitem[Khandelwal et~al\mbox{.}(2019)]%
        {khandelwal2019generalization}
\bibfield{author}{\bibinfo{person}{Urvashi Khandelwal}, \bibinfo{person}{Omer Levy}, \bibinfo{person}{Dan Jurafsky}, \bibinfo{person}{Luke Zettlemoyer}, {and} \bibinfo{person}{Mike Lewis}.} \bibinfo{year}{2019}\natexlab{}.
\newblock \showarticletitle{Generalization through memorization: Nearest neighbor language models}.
\newblock \bibinfo{journal}{\emph{arXiv preprint arXiv:1911.00172}} (\bibinfo{year}{2019}).
\newblock


\bibitem[Kim et~al\mbox{.}(2021)]%
        {kim2021code}
\bibfield{author}{\bibinfo{person}{Seohyun Kim}, \bibinfo{person}{Jinman Zhao}, \bibinfo{person}{Yuchi Tian}, {and} \bibinfo{person}{Satish Chandra}.} \bibinfo{year}{2021}\natexlab{}.
\newblock \showarticletitle{Code prediction by feeding trees to transformers}. In \bibinfo{booktitle}{\emph{2021 IEEE/ACM 43rd International Conference on Software Engineering (ICSE)}}. IEEE, \bibinfo{pages}{150--162}.
\newblock


\bibitem[Lewis et~al\mbox{.}(2020)]%
        {lewis2020retrieval}
\bibfield{author}{\bibinfo{person}{Patrick Lewis}, \bibinfo{person}{Ethan Perez}, \bibinfo{person}{Aleksandra Piktus}, \bibinfo{person}{Fabio Petroni}, \bibinfo{person}{Vladimir Karpukhin}, \bibinfo{person}{Naman Goyal}, \bibinfo{person}{Heinrich K{\"u}ttler}, \bibinfo{person}{Mike Lewis}, \bibinfo{person}{Wen-tau Yih}, \bibinfo{person}{Tim Rockt{\"a}schel}, {et~al\mbox{.}}} \bibinfo{year}{2020}\natexlab{}.
\newblock \showarticletitle{Retrieval-augmented generation for knowledge-intensive nlp tasks}.
\newblock \bibinfo{journal}{\emph{Advances in neural information processing systems}}  \bibinfo{volume}{33} (\bibinfo{year}{2020}), \bibinfo{pages}{9459--9474}.
\newblock


\bibitem[Li et~al\mbox{.}(2017)]%
        {li2017code}
\bibfield{author}{\bibinfo{person}{Jian Li}, \bibinfo{person}{Yue Wang}, \bibinfo{person}{Michael~R Lyu}, {and} \bibinfo{person}{Irwin King}.} \bibinfo{year}{2017}\natexlab{}.
\newblock \showarticletitle{Code completion with neural attention and pointer networks}.
\newblock \bibinfo{journal}{\emph{arXiv preprint arXiv:1711.09573}} (\bibinfo{year}{2017}).
\newblock


\bibitem[Li et~al\mbox{.}(2023b)]%
        {li2023acecoder}
\bibfield{author}{\bibinfo{person}{Jia Li}, \bibinfo{person}{Yunfei Zhao}, \bibinfo{person}{Yongmin Li}, \bibinfo{person}{Ge Li}, {and} \bibinfo{person}{Zhi Jin}.} \bibinfo{year}{2023}\natexlab{b}.
\newblock \showarticletitle{Acecoder: Utilizing existing code to enhance code generation}.
\newblock \bibinfo{journal}{\emph{arXiv preprint arXiv:2303.17780}} (\bibinfo{year}{2023}).
\newblock


\bibitem[Li et~al\mbox{.}(2023a)]%
        {li2023starcoder}
\bibfield{author}{\bibinfo{person}{Raymond Li}, \bibinfo{person}{Loubna~Ben Allal}, \bibinfo{person}{Yangtian Zi}, \bibinfo{person}{Niklas Muennighoff}, \bibinfo{person}{Denis Kocetkov}, \bibinfo{person}{Chenghao Mou}, \bibinfo{person}{Marc Marone}, \bibinfo{person}{Christopher Akiki}, \bibinfo{person}{Jia Li}, \bibinfo{person}{Jenny Chim}, {et~al\mbox{.}}} \bibinfo{year}{2023}\natexlab{a}.
\newblock \showarticletitle{Starcoder: may the source be with you!}
\newblock \bibinfo{journal}{\emph{arXiv preprint arXiv:2305.06161}} (\bibinfo{year}{2023}).
\newblock


\bibitem[Liang et~al\mbox{.}(2024)]%
        {liang2024repofuse}
\bibfield{author}{\bibinfo{person}{Ming Liang}, \bibinfo{person}{Xiaoheng Xie}, \bibinfo{person}{Gehao Zhang}, \bibinfo{person}{Xunjin Zheng}, \bibinfo{person}{Peng Di}, \bibinfo{person}{Hongwei Chen}, \bibinfo{person}{Chengpeng Wang}, \bibinfo{person}{Gang Fan}, {et~al\mbox{.}}} \bibinfo{year}{2024}\natexlab{}.
\newblock \showarticletitle{Repofuse: Repository-level code completion with fused dual context}.
\newblock \bibinfo{journal}{\emph{arXiv preprint arXiv:2402.14323}} (\bibinfo{year}{2024}).
\newblock


\bibitem[Liao et~al\mbox{.}(2023)]%
        {liao2023context}
\bibfield{author}{\bibinfo{person}{Dianshu Liao}, \bibinfo{person}{Shidong Pan}, \bibinfo{person}{Qing Huang}, \bibinfo{person}{Xiaoxue Ren}, \bibinfo{person}{Zhenchang Xing}, \bibinfo{person}{Huan Jin}, {and} \bibinfo{person}{Qinying Li}.} \bibinfo{year}{2023}\natexlab{}.
\newblock \showarticletitle{Context-aware code generation framework for code repositories: Local, global, and third-party library awareness}.
\newblock \bibinfo{journal}{\emph{CoRR}} (\bibinfo{year}{2023}).
\newblock


\bibitem[Liu et~al\mbox{.}(2024a)]%
        {liu2024deepseek}
\bibfield{author}{\bibinfo{person}{Aixin Liu}, \bibinfo{person}{Bei Feng}, \bibinfo{person}{Bing Xue}, \bibinfo{person}{Bingxuan Wang}, \bibinfo{person}{Bochao Wu}, \bibinfo{person}{Chengda Lu}, \bibinfo{person}{Chenggang Zhao}, \bibinfo{person}{Chengqi Deng}, \bibinfo{person}{Chenyu Zhang}, \bibinfo{person}{Chong Ruan}, {et~al\mbox{.}}} \bibinfo{year}{2024}\natexlab{a}.
\newblock \showarticletitle{Deepseek-v3 technical report}.
\newblock \bibinfo{journal}{\emph{arXiv preprint arXiv:2412.19437}} (\bibinfo{year}{2024}).
\newblock


\bibitem[Liu et~al\mbox{.}(2020)]%
        {liu2020multi}
\bibfield{author}{\bibinfo{person}{Fang Liu}, \bibinfo{person}{Ge Li}, \bibinfo{person}{Yunfei Zhao}, {and} \bibinfo{person}{Zhi Jin}.} \bibinfo{year}{2020}\natexlab{}.
\newblock \showarticletitle{Multi-task learning based pre-trained language model for code completion}. In \bibinfo{booktitle}{\emph{Proceedings of the 35th IEEE/ACM international conference on automated software engineering}}. \bibinfo{pages}{473--485}.
\newblock


\bibitem[Liu et~al\mbox{.}(2023a)]%
        {liu2023lost}
\bibfield{author}{\bibinfo{person}{Nelson~F Liu}, \bibinfo{person}{Kevin Lin}, \bibinfo{person}{John Hewitt}, \bibinfo{person}{Ashwin Paranjape}, \bibinfo{person}{Michele Bevilacqua}, \bibinfo{person}{Fabio Petroni}, {and} \bibinfo{person}{Percy Liang}.} \bibinfo{year}{2023}\natexlab{a}.
\newblock \showarticletitle{Lost in the middle: How language models use long contexts}.
\newblock \bibinfo{journal}{\emph{arXiv preprint arXiv:2307.03172}} (\bibinfo{year}{2023}).
\newblock


\bibitem[Liu et~al\mbox{.}(2023b)]%
        {liu2023repobench}
\bibfield{author}{\bibinfo{person}{Tianyang Liu}, \bibinfo{person}{Canwen Xu}, {and} \bibinfo{person}{Julian McAuley}.} \bibinfo{year}{2023}\natexlab{b}.
\newblock \showarticletitle{Repobench: Benchmarking repository-level code auto-completion systems}.
\newblock \bibinfo{journal}{\emph{arXiv preprint arXiv:2306.03091}} (\bibinfo{year}{2023}).
\newblock


\bibitem[Liu et~al\mbox{.}(2024b)]%
        {liu2024graphcoder}
\bibfield{author}{\bibinfo{person}{Wei Liu}, \bibinfo{person}{Ailun Yu}, \bibinfo{person}{Daoguang Zan}, \bibinfo{person}{Bo Shen}, \bibinfo{person}{Wei Zhang}, \bibinfo{person}{Haiyan Zhao}, \bibinfo{person}{Zhi Jin}, {and} \bibinfo{person}{Qianxiang Wang}.} \bibinfo{year}{2024}\natexlab{b}.
\newblock \showarticletitle{Graphcoder: Enhancing repository-level code completion via code context graph-based retrieval and language model}.
\newblock \bibinfo{journal}{\emph{arXiv preprint arXiv:2406.07003}} (\bibinfo{year}{2024}).
\newblock


\bibitem[Lozhkov et~al\mbox{.}(2024)]%
        {lozhkov2024starcoder}
\bibfield{author}{\bibinfo{person}{Anton Lozhkov}, \bibinfo{person}{Raymond Li}, \bibinfo{person}{Loubna~Ben Allal}, \bibinfo{person}{Federico Cassano}, \bibinfo{person}{Joel Lamy-Poirier}, \bibinfo{person}{Nouamane Tazi}, \bibinfo{person}{Ao Tang}, \bibinfo{person}{Dmytro Pykhtar}, \bibinfo{person}{Jiawei Liu}, \bibinfo{person}{Yuxiang Wei}, {et~al\mbox{.}}} \bibinfo{year}{2024}\natexlab{}.
\newblock \showarticletitle{Starcoder 2 and the stack v2: The next generation}.
\newblock \bibinfo{journal}{\emph{arXiv preprint arXiv:2402.19173}} (\bibinfo{year}{2024}).
\newblock


\bibitem[Luo et~al\mbox{.}(2023)]%
        {luo2023wizardcoder}
\bibfield{author}{\bibinfo{person}{Ziyang Luo}, \bibinfo{person}{Can Xu}, \bibinfo{person}{Pu Zhao}, \bibinfo{person}{Qingfeng Sun}, \bibinfo{person}{Xiubo Geng}, \bibinfo{person}{Wenxiang Hu}, \bibinfo{person}{Chongyang Tao}, \bibinfo{person}{Jing Ma}, \bibinfo{person}{Qingwei Lin}, {and} \bibinfo{person}{Daxin Jiang}.} \bibinfo{year}{2023}\natexlab{}.
\newblock \showarticletitle{Wizardcoder: Empowering code large language models with evol-instruct}.
\newblock \bibinfo{journal}{\emph{arXiv preprint arXiv:2306.08568}} (\bibinfo{year}{2023}).
\newblock


\bibitem[Nijkamp et~al\mbox{.}(2023)]%
        {nijkamp2023codegen2}
\bibfield{author}{\bibinfo{person}{Erik Nijkamp}, \bibinfo{person}{Hiroaki Hayashi}, \bibinfo{person}{Caiming Xiong}, \bibinfo{person}{Silvio Savarese}, {and} \bibinfo{person}{Yingbo Zhou}.} \bibinfo{year}{2023}\natexlab{}.
\newblock \showarticletitle{Codegen2: Lessons for training llms on programming and natural languages}.
\newblock \bibinfo{journal}{\emph{arXiv preprint arXiv:2305.02309}} (\bibinfo{year}{2023}).
\newblock


\bibitem[OpenAI(2024)]%
        {openai2024gpt4o}
\bibfield{author}{\bibinfo{person}{OpenAI}.} \bibinfo{year}{2024}\natexlab{}.
\newblock \bibinfo{title}{Hello GPT-4o}.
\newblock \bibinfo{howpublished}{\url{https://openai.com/index/hello-gpt-4o/}}.
\newblock
\newblock
\shownote{2024}.


\bibitem[Parvez et~al\mbox{.}(2021)]%
        {parvez2021retrieval}
\bibfield{author}{\bibinfo{person}{Md~Rizwan Parvez}, \bibinfo{person}{Wasi~Uddin Ahmad}, \bibinfo{person}{Saikat Chakraborty}, \bibinfo{person}{Baishakhi Ray}, {and} \bibinfo{person}{Kai-Wei Chang}.} \bibinfo{year}{2021}\natexlab{}.
\newblock \showarticletitle{Retrieval augmented code generation and summarization}.
\newblock \bibinfo{journal}{\emph{arXiv preprint arXiv:2108.11601}} (\bibinfo{year}{2021}).
\newblock


\bibitem[Peng et~al\mbox{.}(2024)]%
        {peng2024graph}
\bibfield{author}{\bibinfo{person}{Boci Peng}, \bibinfo{person}{Yun Zhu}, \bibinfo{person}{Yongchao Liu}, \bibinfo{person}{Xiaohe Bo}, \bibinfo{person}{Haizhou Shi}, \bibinfo{person}{Chuntao Hong}, \bibinfo{person}{Yan Zhang}, {and} \bibinfo{person}{Siliang Tang}.} \bibinfo{year}{2024}\natexlab{}.
\newblock \showarticletitle{Graph retrieval-augmented generation: A survey}.
\newblock \bibinfo{journal}{\emph{arXiv preprint arXiv:2408.08921}} (\bibinfo{year}{2024}).
\newblock


\bibitem[Phan et~al\mbox{.}(2024)]%
        {phan2024repohyper}
\bibfield{author}{\bibinfo{person}{Huy~Nhat Phan}, \bibinfo{person}{Hoang~Nhat Phan}, \bibinfo{person}{Tien~N Nguyen}, {and} \bibinfo{person}{Nghi~DQ Bui}.} \bibinfo{year}{2024}\natexlab{}.
\newblock \showarticletitle{Repohyper: Better context retrieval is all you need for repository-level code completion}.
\newblock \bibinfo{journal}{\emph{CoRR}} (\bibinfo{year}{2024}).
\newblock


\bibitem[Qwen et~al\mbox{.}(2025)]%
        {qwen2025qwen25technicalreport}
\bibfield{author}{\bibinfo{person}{Qwen}, \bibinfo{person}{:}, \bibinfo{person}{An Yang}, \bibinfo{person}{Baosong Yang}, \bibinfo{person}{Beichen Zhang}, \bibinfo{person}{Binyuan Hui}, \bibinfo{person}{Bo Zheng}, \bibinfo{person}{Bowen Yu}, \bibinfo{person}{Chengyuan Li}, \bibinfo{person}{Dayiheng Liu}, \bibinfo{person}{Fei Huang}, \bibinfo{person}{Haoran Wei}, \bibinfo{person}{Huan Lin}, \bibinfo{person}{Jian Yang}, \bibinfo{person}{Jianhong Tu}, \bibinfo{person}{Jianwei Zhang}, \bibinfo{person}{Jianxin Yang}, \bibinfo{person}{Jiaxi Yang}, \bibinfo{person}{Jingren Zhou}, \bibinfo{person}{Junyang Lin}, \bibinfo{person}{Kai Dang}, \bibinfo{person}{Keming Lu}, \bibinfo{person}{Keqin Bao}, \bibinfo{person}{Kexin Yang}, \bibinfo{person}{Le Yu}, \bibinfo{person}{Mei Li}, \bibinfo{person}{Mingfeng Xue}, \bibinfo{person}{Pei Zhang}, \bibinfo{person}{Qin Zhu}, \bibinfo{person}{Rui Men}, \bibinfo{person}{Runji Lin}, \bibinfo{person}{Tianhao Li}, \bibinfo{person}{Tianyi Tang}, \bibinfo{person}{Tingyu Xia}, \bibinfo{person}{Xingzhang Ren}, \bibinfo{person}{Xuancheng Ren}, \bibinfo{person}{Yang Fan}, \bibinfo{person}{Yang Su}, \bibinfo{person}{Yichang Zhang}, \bibinfo{person}{Yu Wan}, \bibinfo{person}{Yuqiong Liu}, \bibinfo{person}{Zeyu Cui}, \bibinfo{person}{Zhenru Zhang}, {and} \bibinfo{person}{Zihan Qiu}.} \bibinfo{year}{2025}\natexlab{}.
\newblock \bibinfo{title}{Qwen2.5 Technical Report}.
\newblock
\showeprint[arxiv]{2412.15115}~[cs.CL]
\urldef\tempurl%
\url{https://arxiv.org/abs/2412.15115}
\showURL{%
\tempurl}


\bibitem[Robertson et~al\mbox{.}(2009)]%
        {robertson2009probabilistic}
\bibfield{author}{\bibinfo{person}{Stephen Robertson}, \bibinfo{person}{Hugo Zaragoza}, {et~al\mbox{.}}} \bibinfo{year}{2009}\natexlab{}.
\newblock \showarticletitle{The probabilistic relevance framework: BM25 and beyond}.
\newblock \bibinfo{journal}{\emph{Foundations and Trends{\textregistered} in Information Retrieval}} \bibinfo{volume}{3}, \bibinfo{number}{4} (\bibinfo{year}{2009}), \bibinfo{pages}{333--389}.
\newblock


\bibitem[Roziere et~al\mbox{.}(2023)]%
        {roziere2023code}
\bibfield{author}{\bibinfo{person}{Baptiste Roziere}, \bibinfo{person}{Jonas Gehring}, \bibinfo{person}{Fabian Gloeckle}, \bibinfo{person}{Sten Sootla}, \bibinfo{person}{Itai Gat}, \bibinfo{person}{Xiaoqing~Ellen Tan}, \bibinfo{person}{Yossi Adi}, \bibinfo{person}{Jingyu Liu}, \bibinfo{person}{Romain Sauvestre}, \bibinfo{person}{Tal Remez}, {et~al\mbox{.}}} \bibinfo{year}{2023}\natexlab{}.
\newblock \showarticletitle{Code llama: Open foundation models for code}.
\newblock \bibinfo{journal}{\emph{arXiv preprint arXiv:2308.12950}} (\bibinfo{year}{2023}).
\newblock


\bibitem[Shi et~al\mbox{.}(2023)]%
        {shi2023large}
\bibfield{author}{\bibinfo{person}{Freda Shi}, \bibinfo{person}{Xinyun Chen}, \bibinfo{person}{Kanishka Misra}, \bibinfo{person}{Nathan Scales}, \bibinfo{person}{David Dohan}, \bibinfo{person}{Ed~H Chi}, \bibinfo{person}{Nathanael Sch{\"a}rli}, {and} \bibinfo{person}{Denny Zhou}.} \bibinfo{year}{2023}\natexlab{}.
\newblock \showarticletitle{Large language models can be easily distracted by irrelevant context}. In \bibinfo{booktitle}{\emph{International Conference on Machine Learning}}. PMLR, \bibinfo{pages}{31210--31227}.
\newblock


\bibitem[Shrivastava et~al\mbox{.}(2023a)]%
        {shrivastava2023repofusion}
\bibfield{author}{\bibinfo{person}{Disha Shrivastava}, \bibinfo{person}{Denis Kocetkov}, \bibinfo{person}{Harm de Vries}, \bibinfo{person}{Dzmitry Bahdanau}, {and} \bibinfo{person}{Torsten Scholak}.} \bibinfo{year}{2023}\natexlab{a}.
\newblock \showarticletitle{Repofusion: Training code models to understand your repository}.
\newblock \bibinfo{journal}{\emph{arXiv preprint arXiv:2306.10998}} (\bibinfo{year}{2023}).
\newblock


\bibitem[Shrivastava et~al\mbox{.}(2023b)]%
        {shrivastava2023repository}
\bibfield{author}{\bibinfo{person}{Disha Shrivastava}, \bibinfo{person}{Hugo Larochelle}, {and} \bibinfo{person}{Daniel Tarlow}.} \bibinfo{year}{2023}\natexlab{b}.
\newblock \showarticletitle{Repository-level prompt generation for large language models of code}. In \bibinfo{booktitle}{\emph{International Conference on Machine Learning}}. PMLR, \bibinfo{pages}{31693--31715}.
\newblock


\bibitem[Tan et~al\mbox{.}(2024)]%
        {tan2024prompt}
\bibfield{author}{\bibinfo{person}{Hanzhuo Tan}, \bibinfo{person}{Qi Luo}, \bibinfo{person}{Ling Jiang}, \bibinfo{person}{Zizheng Zhan}, \bibinfo{person}{Jing Li}, \bibinfo{person}{Haotian Zhang}, {and} \bibinfo{person}{Yuqun Zhang}.} \bibinfo{year}{2024}\natexlab{}.
\newblock \showarticletitle{Prompt-based code completion via multi-retrieval augmented generation}.
\newblock \bibinfo{journal}{\emph{ACM Transactions on Software Engineering and Methodology}} (\bibinfo{year}{2024}).
\newblock


\bibitem[Tang et~al\mbox{.}(2023)]%
        {tang2023domain}
\bibfield{author}{\bibinfo{person}{Ze Tang}, \bibinfo{person}{Jidong Ge}, \bibinfo{person}{Shangqing Liu}, \bibinfo{person}{Tingwei Zhu}, \bibinfo{person}{Tongtong Xu}, \bibinfo{person}{Liguo Huang}, {and} \bibinfo{person}{Bin Luo}.} \bibinfo{year}{2023}\natexlab{}.
\newblock \showarticletitle{Domain adaptive code completion via language models and decoupled domain databases}. In \bibinfo{booktitle}{\emph{2023 38th IEEE/ACM International Conference on Automated Software Engineering (ASE)}}. IEEE, \bibinfo{pages}{421--433}.
\newblock


\bibitem[Tao et~al\mbox{.}(2025)]%
        {tao2025code}
\bibfield{author}{\bibinfo{person}{Hongyuan Tao}, \bibinfo{person}{Ying Zhang}, \bibinfo{person}{Zhenhao Tang}, \bibinfo{person}{Hongen Peng}, \bibinfo{person}{Xukun Zhu}, \bibinfo{person}{Bingchang Liu}, \bibinfo{person}{Yingguang Yang}, \bibinfo{person}{Ziyin Zhang}, \bibinfo{person}{Zhaogui Xu}, \bibinfo{person}{Haipeng Zhang}, {et~al\mbox{.}}} \bibinfo{year}{2025}\natexlab{}.
\newblock \showarticletitle{Code Graph Model (CGM): A Graph-Integrated Large Language Model for Repository-Level Software Engineering Tasks}.
\newblock \bibinfo{journal}{\emph{arXiv preprint arXiv:2505.16901}} (\bibinfo{year}{2025}).
\newblock


\bibitem[Wang et~al\mbox{.}(2023)]%
        {wang2023codet5+}
\bibfield{author}{\bibinfo{person}{Yue Wang}, \bibinfo{person}{Hung Le}, \bibinfo{person}{Akhilesh~Deepak Gotmare}, \bibinfo{person}{Nghi~DQ Bui}, \bibinfo{person}{Junnan Li}, {and} \bibinfo{person}{Steven~CH Hoi}.} \bibinfo{year}{2023}\natexlab{}.
\newblock \showarticletitle{Codet5+: Open code large language models for code understanding and generation}.
\newblock \bibinfo{journal}{\emph{arXiv preprint arXiv:2305.07922}} (\bibinfo{year}{2023}).
\newblock


\bibitem[Wang et~al\mbox{.}(2024)]%
        {wang2024rlcoder}
\bibfield{author}{\bibinfo{person}{Yanlin Wang}, \bibinfo{person}{Yanli Wang}, \bibinfo{person}{Daya Guo}, \bibinfo{person}{Jiachi Chen}, \bibinfo{person}{Ruikai Zhang}, \bibinfo{person}{Yuchi Ma}, {and} \bibinfo{person}{Zibin Zheng}.} \bibinfo{year}{2024}\natexlab{}.
\newblock \showarticletitle{Rlcoder: Reinforcement learning for repository-level code completion}.
\newblock \bibinfo{journal}{\emph{arXiv preprint arXiv:2407.19487}} (\bibinfo{year}{2024}).
\newblock


\bibitem[Yadavalli et~al\mbox{.}(2023)]%
        {yadavalli2023rlpg}
\bibfield{author}{\bibinfo{person}{Sushma~Reddy Yadavalli}, \bibinfo{person}{Lokesh~Chandra Das}, {and} \bibinfo{person}{Myounggyu Won}.} \bibinfo{year}{2023}\natexlab{}.
\newblock \showarticletitle{Rlpg: Reinforcement learning approach for dynamic intra-platoon gap adaptation for highway on-ramp merging}. In \bibinfo{booktitle}{\emph{2023 IEEE/RSJ International Conference on Intelligent Robots and Systems (IROS)}}. IEEE, \bibinfo{pages}{5514--5521}.
\newblock


\bibitem[Zan et~al\mbox{.}(2022a)]%
        {zan2022language}
\bibfield{author}{\bibinfo{person}{Daoguang Zan}, \bibinfo{person}{Bei Chen}, \bibinfo{person}{Zeqi Lin}, \bibinfo{person}{Bei Guan}, \bibinfo{person}{Yongji Wang}, {and} \bibinfo{person}{Jian-Guang Lou}.} \bibinfo{year}{2022}\natexlab{a}.
\newblock \showarticletitle{When language model meets private library}.
\newblock \bibinfo{journal}{\emph{arXiv preprint arXiv:2210.17236}} (\bibinfo{year}{2022}).
\newblock


\bibitem[Zan et~al\mbox{.}(2022b)]%
        {zan2022cert}
\bibfield{author}{\bibinfo{person}{Daoguang Zan}, \bibinfo{person}{Bei Chen}, \bibinfo{person}{Dejian Yang}, \bibinfo{person}{Zeqi Lin}, \bibinfo{person}{Minsu Kim}, \bibinfo{person}{Bei Guan}, \bibinfo{person}{Yongji Wang}, \bibinfo{person}{Weizhu Chen}, {and} \bibinfo{person}{Jian-Guang Lou}.} \bibinfo{year}{2022}\natexlab{b}.
\newblock \showarticletitle{CERT: continual pre-training on sketches for library-oriented code generation}.
\newblock \bibinfo{journal}{\emph{arXiv preprint arXiv:2206.06888}} (\bibinfo{year}{2022}).
\newblock


\bibitem[Zan et~al\mbox{.}(2022c)]%
        {zan2022large}
\bibfield{author}{\bibinfo{person}{Daoguang Zan}, \bibinfo{person}{Bei Chen}, \bibinfo{person}{Fengji Zhang}, \bibinfo{person}{Dianjie Lu}, \bibinfo{person}{Bingchao Wu}, \bibinfo{person}{Bei Guan}, \bibinfo{person}{Yongji Wang}, {and} \bibinfo{person}{Jian-Guang Lou}.} \bibinfo{year}{2022}\natexlab{c}.
\newblock \showarticletitle{Large language models meet NL2Code: A survey}.
\newblock \bibinfo{journal}{\emph{arXiv preprint arXiv:2212.09420}} (\bibinfo{year}{2022}).
\newblock


\bibitem[Zhang et~al\mbox{.}(2023b)]%
        {zhang2023repocoder}
\bibfield{author}{\bibinfo{person}{Fengji Zhang}, \bibinfo{person}{Bei Chen}, \bibinfo{person}{Yue Zhang}, \bibinfo{person}{Jacky Keung}, \bibinfo{person}{Jin Liu}, \bibinfo{person}{Daoguang Zan}, \bibinfo{person}{Yi Mao}, \bibinfo{person}{Jian-Guang Lou}, {and} \bibinfo{person}{Weizhu Chen}.} \bibinfo{year}{2023}\natexlab{b}.
\newblock \showarticletitle{Repocoder: Repository-level code completion through iterative retrieval and generation}.
\newblock \bibinfo{journal}{\emph{arXiv preprint arXiv:2303.12570}} (\bibinfo{year}{2023}).
\newblock


\bibitem[Zhang et~al\mbox{.}(2023c)]%
        {zhang2023syntax}
\bibfield{author}{\bibinfo{person}{Xiangyu Zhang}, \bibinfo{person}{Yu Zhou}, \bibinfo{person}{Guang Yang}, {and} \bibinfo{person}{Taolue Chen}.} \bibinfo{year}{2023}\natexlab{c}.
\newblock \showarticletitle{Syntax-aware retrieval augmented code generation}. In \bibinfo{booktitle}{\emph{Findings of the Association for Computational Linguistics: EMNLP 2023}}. \bibinfo{pages}{1291--1302}.
\newblock


\bibitem[Zhang et~al\mbox{.}(2023a)]%
        {zhang2023unifying}
\bibfield{author}{\bibinfo{person}{Ziyin Zhang}, \bibinfo{person}{Chaoyu Chen}, \bibinfo{person}{Bingchang Liu}, \bibinfo{person}{Cong Liao}, \bibinfo{person}{Zi Gong}, \bibinfo{person}{Hang Yu}, \bibinfo{person}{Jianguo Li}, {and} \bibinfo{person}{Rui Wang}.} \bibinfo{year}{2023}\natexlab{a}.
\newblock \showarticletitle{Unifying the perspectives of nlp and software engineering: A survey on language models for code}.
\newblock \bibinfo{journal}{\emph{arXiv preprint arXiv:2311.07989}} (\bibinfo{year}{2023}).
\newblock


\bibitem[Zheng et~al\mbox{.}(2023b)]%
        {zheng2023codegeex}
\bibfield{author}{\bibinfo{person}{Qinkai Zheng}, \bibinfo{person}{Xiao Xia}, \bibinfo{person}{Xu Zou}, \bibinfo{person}{Yuxiao Dong}, \bibinfo{person}{Shan Wang}, \bibinfo{person}{Yufei Xue}, \bibinfo{person}{Lei Shen}, \bibinfo{person}{Zihan Wang}, \bibinfo{person}{Andi Wang}, \bibinfo{person}{Yang Li}, {et~al\mbox{.}}} \bibinfo{year}{2023}\natexlab{b}.
\newblock \showarticletitle{Codegeex: A pre-trained model for code generation with multilingual benchmarking on humaneval-x}. In \bibinfo{booktitle}{\emph{Proceedings of the 29th ACM SIGKDD Conference on Knowledge Discovery and Data Mining}}. \bibinfo{pages}{5673--5684}.
\newblock


\bibitem[Zheng et~al\mbox{.}(2023a)]%
        {zheng2023survey}
\bibfield{author}{\bibinfo{person}{Zibin Zheng}, \bibinfo{person}{Kaiwen Ning}, \bibinfo{person}{Yanlin Wang}, \bibinfo{person}{Jingwen Zhang}, \bibinfo{person}{Dewu Zheng}, \bibinfo{person}{Mingxi Ye}, {and} \bibinfo{person}{Jiachi Chen}.} \bibinfo{year}{2023}\natexlab{a}.
\newblock \showarticletitle{A survey of large language models for code: Evolution, benchmarking, and future trends}.
\newblock \bibinfo{journal}{\emph{arXiv preprint arXiv:2311.10372}} (\bibinfo{year}{2023}).
\newblock


\bibitem[Zhu et~al\mbox{.}(2024)]%
        {zhu2024deepseek}
\bibfield{author}{\bibinfo{person}{Qihao Zhu}, \bibinfo{person}{Daya Guo}, \bibinfo{person}{Zhihong Shao}, \bibinfo{person}{Dejian Yang}, \bibinfo{person}{Peiyi Wang}, \bibinfo{person}{Runxin Xu}, \bibinfo{person}{Y Wu}, \bibinfo{person}{Yukun Li}, \bibinfo{person}{Huazuo Gao}, \bibinfo{person}{Shirong Ma}, {et~al\mbox{.}}} \bibinfo{year}{2024}\natexlab{}.
\newblock \showarticletitle{Deepseek-coder-v2: Breaking the barrier of closed-source models in code intelligence}.
\newblock \bibinfo{journal}{\emph{arXiv preprint arXiv:2406.11931}} (\bibinfo{year}{2024}).
\newblock


\end{thebibliography}


\end{document}